\pdfoutput=1
\documentclass[twocolumn,a4paper,amsmath,amsfonts,10pt,nopdfoutputerror,accepted=2022-02-25]{quantumarticle}

\usepackage[T1]{fontenc}
\usepackage[colorlinks, linkcolor=linkblue]{hyperref}
\let\doi\relax
\usepackage[caption=false, labelfont=bf]{subfig}
\DeclareCaptionJustification{justified}{\justifying}
\usepackage{aliascnt,cleveref,booktabs,microtype,hyphenat,ragged2e,xcolor,tikz,pgfplots,stmaryrd,mathtools,braket,bm,array,amsthm,url,doi,afterpage,float,enumitem}
\usepackage[numbers,sort&compress,merge]{natbib}
\usepackage[compat=0.4]{yquant}

\usetikzlibrary{tikzmark,decorations.pathreplacing,calc,fpu,decorations.pathmorphing,arrows.meta}
\pgfplotsset{compat=1.16, every axis plot/.append style={thick}, filter discard warning=false}
\definecolor{mpc0}        {RGB} {94, 129, 181}
\definecolor{mpc1}        {RGB} {225, 156, 36}
\definecolor{mpc2}        {RGB} {143, 176, 50}
\definecolor{mpc3}        {RGB} {235, 98, 53}
\definecolor{mpc4}        {RGB} {135, 120, 179}
\definecolor{mpc5}        {RGB} {197, 110, 26}
\definecolor{mpc6}        {RGB} {93, 158, 199}
\definecolor{mpc7}        {RGB} {255, 191, 0}
\definecolor{mpc8}        {RGB} {165, 96, 157}
\definecolor{mpc9}        {RGB} {146, 150, 0}

\definecolor{darkred}     {RGB} {166,  10,  18}
\definecolor{darkgreen}   {RGB} { 15, 112,  20}
\definecolor{darkblue}    {RGB} {  5,  10, 122}
\definecolor{bob}         {RGB} {16,   97, 157}
\definecolor{alice}       {RGB} {190,  29,  44}
\colorlet{linkblue}       {cyan!35!blue}

\crefname{pluralequation}{equations}{equations}
\crefname{program}{program}{programs}
\crefname{programs}{programs}{programs}

\makeatletter
\tikzset{font append/.style={font/.expand once=\tikz@textfont #1},
         font append/.value required}

\protected\def\subblock@finish{%
   \endgroup%
   \global\pgflinewidth=\tikzscope@linewidth\relax%
   \tikz@path@do@at@end%
}

\newcommand\algoblock[3][]{%
   \begingroup%
      \path (lastnode.south west) ++(0, -\boxsep) coordinate (lastnode);%
      \pgfpointanchor{lastnode}{south west}%
      \edef\storedlastnode{\noexpand\coordinate (lastnode) at (\the\pgf@x, \the\pgf@y);}%
      \edef\content{%
         \ifstrempty{#1}{}{%
            \unexpanded{%
               \path let \p1=(lastnode.south west) in%
                  (.5*\x1+.5*\colright-.5*\xsep, \y1) ++(0, -\boxsep)
                  node[anchor=north, inner sep=0pt, font append=\ttfamily] (lastnode) {#1};
            }%
         }%
         \noexpand\path let \noexpand\p1=(lastnode.south west) in
            (\the\dimexpr\pgf@x+\xsep\relax, \noexpand\y1) coordinate (lastnode);%
         \unexpanded{#3}%
         \noexpand\path[overlay] let \noexpand\p1=(lastnode.south west) in
            (\the\pgf@x, \noexpand\y1) ++(0, -\boxsep) coordinate (lastnode);%
      }%
      \let\path=\tikz@command@path%
      \let\tikz@finish=\subblock@finish%
      \pgfinterruptboundingbox%
         \begingroup%
            \content%
         \endgroup%
         \edef\getmeasure{%
               \noexpand\endpgfinterruptboundingbox%
            \endgroup%
            \unexpanded\expandafter{\storedlastnode}% nodes are set globally
            \noexpand\path[{#2}] let \noexpand\p1=(lastnode.south west) in%
               (\colright, \noexpand\y1-\the\dimexpr\pgf@picmaxy-\pgf@picminy\relax)
               ++(0, -\boxsep) rectangle (\noexpand\p1);%
            \unexpanded\expandafter{\content}%
         }%
   \getmeasure%
}

\newcommand\algosubblock[3][]{%
   \path[overlay] (lastnode.south west) ++(\indentsep, 0) coordinate (lastnode);%
   \algoblock[{#1}]{#2}{#3}%
   \path[overlay] (lastnode.south west) ++(-\indentsep, 0) coordinate (lastnode);%
}

\newcommand\algosub[1]{%
   \path[overlay] (lastnode.south west) ++(\indentsep, 0) coordinate (lastnode);%
   #1%
   \path[overlay] (lastnode.south west) ++(-\indentsep, 0) coordinate (lastnode);%
}

\newcommand\algocommand[2][]{%
   \path let \p1=(lastnode.south west) in (\p1) ++(0, -\boxsep)
      node[anchor=north west, inner sep=0pt, #1] (lastnode) {%
         \parbox{\dimexpr\colright-\x1-\xsep\relax}{#2}%
      };%
}

\newenvironment{tikzalgorithm}[1][]{%
   \begin{tikzpicture}[{#1}]
      \coordinate[overlay] (lastnode) at (0, 0);%
      \edef\colright{\the\linewidth}%
      \edef\xsep{1mm}%
      \def\boxsep{.1}%
      \def\indentsep{.5}%
}{\end{tikzpicture}}
\def\makeWithSize#1#2#3{%
   \@namedef{#1}{%
      \begingroup%
      \def\@tmp####1####2{\mathopen{\scale#2}####2\mathclose{\scale#3}\endgroup}%
      \def\middle####1{\mathinner{\scale####1}}%
      \futurelet\scale\@tmp%
   }%
}
\def\makeWithSizeTwo#1#2{%
   \@namedef{#1}{%
      \begingroup%
      \def\@tmp####1####2####3{%
         \def\@argOne{####2}%
         \def\@argTwo{####3}%
         #2%
         \endgroup%
      }%
      \futurelet\scale\@tmp%
   }%
}

\newcommand*\e{{\mathrm e}}
\let\dotlessI=\i
\def\i{\TextOrMath\dotlessI{\mathrm i}}
\renewcommand*\d{\mathop{}\!\mathrm d\mathchoice{}{}{\kern-.09em}{\kern-.09em}}
\renewcommand*\u[2]{#1\,\mathrm{#2}}
\newcommand*\uf[3]{#1\,\mathrm{\frac{#2}{#3}}}
\newcommand*\code[1]{\mathord{\llbracket#1\rrbracket}}

\newcommand*\mat[1]{\begin{pmatrix} #1 \end{pmatrix}}
\newcommand*\ketbra[2]{\ket{#1}\!\bra{#2}}
\newcommand*\abs[1]{\mathinner{\lvert#1\rvert}}
\makeWithSize{Abs}{\lvert}{\rvert}
\newcommand*\norm[1]{\mathinner{\lVert#1\rVert}}
\makeWithSize{Norm}{\lVert}{\rVert}
\DeclareMathOperator\tr{tr}
\DeclareMathOperator\rk{rk}

\let\vv=\bm

\def\compose{\mathbin\circ}
\def\intersect{\mathbin\cap}

\newcommand*\N{\mathbb N}
\newcommand*\Z{\mathbb Z}

\newcommand*\C{\mathbb C}

\newlength{\centerasiflen}
\newcommand*\centerasif[2]{%
   \settowidth{\centerasiflen}{#1}%
   \hskip.5\centerasiflen%
   \mathclap{#2}%
}

\renewcommand*\env@matrix[1][*\c@MaxMatrixCols c]{%
   \hskip -\arraycolsep%
   \let\@ifnextchar\new@ifnextchar%
   \array{#1}%
}
\def\mbig{\bBigg@{1.2}}
\def\mbigl#1{\mathopen{\bBigg@{1.2}#1}}
\def\mbigr#1{\mathclose{\bBigg@{1.2}#1}}

\theoremstyle{definition}

\newtheorem*{definition*}{Definition}

\newaliascnt{theorem}{definition}

\aliascntresetthe{theorem}
\newtheorem*{theorem*}{Theorem}

\newaliascnt{lemma}{definition}

\aliascntresetthe{lemma}
\newtheorem*{lemma*}{Lemma}

\newaliascnt{corollary}{definition}

\aliascntresetthe{corollary}
\newtheorem*{corollary*}{Corollary}

\newaliascnt{remark}{definition}

\aliascntresetthe{remark}
\newtheorem*{remark*}{Remark}

\crefname{definition}{definition}{definitions}
\crefname{theorem}{theorem}{theorems}
\crefname{lemma}{lemma}{lemmas}
\crefname{corollary}{corollary}{corollaries}
\crefname{remark}{remark}{remarks}

\AtBeginDocument{%
   \let\oldcite\@cite%
   \renewcommand\@cite[3]{\mbox{\begingroup\oldcite{#1}{#2}{#3}}\endgroup}%
}

\primitive@output{%
   \dispatch@output
   \global\setbox255=\box\voidb@x%
}%

\appto\@printauthors{\vspace*{-.5em}}

\begin{document}
   \frenchspacing
   \title{Optimizing quantum codes with an application to the loss channel with partial erasure information}
   \author{Benjamin Desef}
   \orcid{0000-0003-2083-7820}
   \author{Martin B. Plenio}
   \orcid{0000-0003-4238-8843}
   \affiliation{Institute for Theoretical Physics \& IQST, University of Ulm}
   \date{2022-02-25}

   \begin{abstract}
      % arXiv pdflatex does provide a good and economic hyphenation experience, which scrambles the layout of the first page. We do it manually.
      \nohyphens{Quantum~error~correcting~codes~(QECCs)
      are~the~means~of~choice~whenever~quantum
      systems~suffer~errors,~e.g.,~due~to~imperfect~de-
      vices,~environments,~or~faulty~channels.~By~now,
      a~plethora~of~families~of~codes~is~known,~but
      there~is~no~universal~approach~to~finding~new~or
      optimal~codes~for~a~certain~task~and~subject~to
      specific~experimental~constraints.~In~particular,
      once~found,~a~QECC~is~typically~used~in~very
      diverse~contexts,~while~its~resilience~against~er-
      rors~is~captured~in~a~single~figure~of~merit,~the
      distance~of~the~code.~This~does~not~necessarily
      give~rise~to~the~most~efficient~protection~possi-
      ble~given~a~certain~known~error~or~a~particular
      application~for~which~the~code~is~employed.

      In~this~paper,~we~investigate~the~loss~chan-
      nel,~which~plays~a~key~role~in~quantum~com-
      munication,~and~in~particular~in~quantum~key
      distribution~over~long~distances.~We~develop~a
      numerical~set~of~tools~that~allows~to~optimize~an
      encoding~specifically~for~recovering~lost~parti-
      cles~both~deterministically~and~probabilistically,
      where~some~knowledge~about~\emph{what}~was~lost~is
      available,~and~demonstrate~its~capabilities.~This
      allows~us~to~arrive~at~new~codes~ideal~for~the
      distribution~of~entangled~states~in~this~particu-
      lar~setting,~and~also~to~investigate~if~encoding~in
      qudits~or~allowing~for~non-deterministic~correc-
      tion~proves~advantageous~compared~to~known
      QECCs.~While~we~here~focus~on~the~case~of
      losses,~our~methodology~is~applicable~whenever
      the~errors~in~a~system~can~be~characterized~by
      a~known~linear~map.}
   \end{abstract}

   \maketitle

   \section{Introduction}
      Quantum key distribution (QKD) uses intrinsic properties of quantum mechanics to allow for information\hyp theoretically secure transmission of information~\cite{bib:Bennett1984,*bib:Bennett1984r,bib:Shor2000,bib:Renner2005}.
      Necessarily, this requires to transmit quantum systems---mostly, photons---between two parties with a sufficiently low error rate~\cite{bib:Scarani2009}.
      Due to the fragility of such systems, a major task in the development of a QKD landscape is to reduce the effects of natural, i.e., not eavesdropper\hyp induced, errors.

      Probably the main issue at the present time is to deal with channel attenuation, which leads to a probabilistic loss of particles~\cite{bib:Agrawal2010,bib:Scarani2009}.
      The exponential nature of channel attenuation makes it practically impossible to develop QKD systems over longer distances with direct transmission~\cite{bib:Takeoka2014,bib:Pirandola2017}.
      Two proposals exist to deal with this issue: quantum repeaters~\cite{bib:Briegel1998,bib:Sangouard2011,bib:Azuma2015,bib:Azuma2016,bib:Pirandola2019} and trusted nodes~\cite{bib:Peev2009,bib:Sasaki2011,bib:Stucki2011}.
      Only the former is able to provide satisfactory security, as the repeater nodes do not have to be trusted.

      In \cref{sec:att}, we will describe the problem of channel attenuation more in detail, with a particular emphasis on QKD.
      We will explain why approaches to and realizations of quantum repeaters currently found in the literature may be valid working examples, but do not address issues that are central to embedding them seamlessly into the current telecom infrastructure.
      This will serve as a motivation in \cref{sec:mult} to introduce a new methodology that has the potential to showcase particular implementations that push the bounds of realistic QKD systems toward their fundamental limits, while conversely, it can also reveal to which extent the infrastructure must be adapted.
      We will apply those methods to various configurations and give results in \cref{sec:num}.
      In \cref{sec:disc}, we will discuss the implications of our results and issues that remain to be addressed.
      We summarize our ideas in \cref{sec:out} and provide an outlook on future potential applications.

   \section{Channel attenuation}\label{sec:att}
      \subsection{Definition and parameter values}
         The secret key rate (per channel use) in QKD is based mainly on two factors: the success probability of transferring the signal, $p_{\mathrm{trans}}$, as well as the quality of the arrived bits, measured in terms of the bit error rates in the chosen basis.
         We employ a simple key rate analysis in the asymptotic framework~\cite{bib:Shor2000,bib:Lo2005}, based on the bit error rates $e_X$ and $e_Z$ in the Pauli~$X$ and~$Z$ bases, respectively, although with a more complicated analysis, improved bounds can be found~\cite{bib:Kraus2005,bib:Renner2005b,bib:Renner2005,bib:Smith2008}.
         We will also disregard any finite size or side channel effects.
         The efficient QKD protocol~\cite{bib:Lo2005} asymptotically removes the sifting prefactor and gives a secret key rate of
         \begin{equation}
            \bigl[ 1 - 2 H(\max\{ e_X, e_Z \}) \bigr] p_{\mathrm{trans}}\text,
            \label{eqn:qkdrate}
         \end{equation}
         where $H$ is the binary entropy function.

         This shows that indeed attenuation---(heralded) loss of signals---does not \emph{fundamentally} limit transmission, but it still does so \emph{effectively}.
         The probability $p_{\mathrm{trans}}$ of successful transmission through a channel of length~$L$ with attenuation coefficient~$\alpha$ is given by
         \begin{equation}
            p_{\mathrm{trans}} = \e^{-\alpha L}\text, \label{eqn:attenuation}
         \end{equation}
         which severely limits the achievable bit rate to drop at least exponentially with distance, and therefore also constrains the distance between repeater stations.

         The ultimate goal of rolling out QKD systems for private and commercial use is to embed them most seamlessly in the \emph{existing} telecom infrastructure.
         While this will require significant investments into new devices such as quantum repeaters and routers, ideally, existing fiber networks should be able to provide the backbone of the QKD network.
         This is not a hard requirement on the way to build a quantum internet, but large\hyp scale infrastructural projects will have a significant impact both on how well the technology will be received in the general public as well as on whether private telecommunication companies will feel inclined to carry out said projects at all.

         Hence, it is crucial to investigate QKD ``at the limit,'' namely, letting both $\alpha$ and $L$ be dictated by the specifications of today's telecommunication networks.
         Common fibers in the $\u{1550}{nm}$ telecom window have an attenuation coefficient $\alpha = \uf{0.2}{dB}{km} = \uf{0.046}{1}{km}$~\cite{bib:Scarani2009}.
         The inter\hyp repeater distance~$L$ is given by the distance between the repeater stations that amplify the classical signals, where unused slots of dark fibers may be employed to plug in quantum repeaters.
         They are located about $\u{60\,\text{--}\,80}{km}$ apart~\cite{bib:Agrawal2010}.

\begin{figure}[htbp]
            \begin{tikzpicture}
               \begin{axis}[xlabel=Distance $L$ ($\mathrm{km}$),
                            enlarge x limits=false,
                            ylabel=$p_{\mathrm{trans}}$, axis on top]
                  \addplot[domain=10:90, samples=100, darkblue] {exp(-0.046*x)};
                  \path
                     (60, 0.0632918) node[pin=100:$\u{6.3}{\%}$, inner sep=0pt] {}
                     (80, 0.025223) node[pin=100:$\u{2.5}{\%}$, inner sep=0pt] {};
               \end{axis}
            \end{tikzpicture}
            \caption{%
               Transmission probability for fiber with attenuation coefficient $\alpha = \uf{0.2}{dB}{km}$%
            }
            \label{fig:ptrans}
         \end{figure}
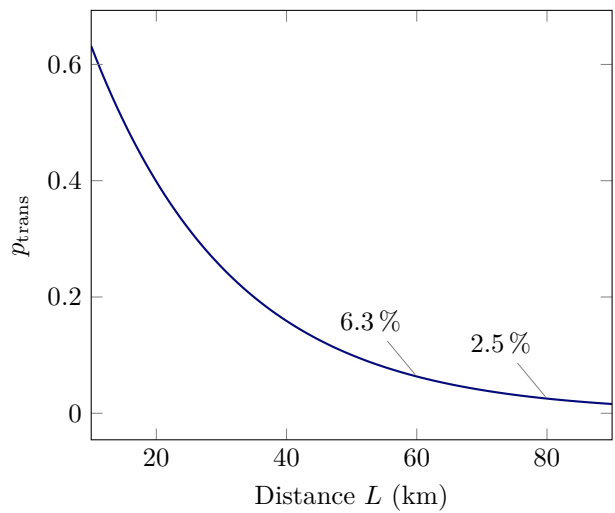

         \Cref{fig:ptrans} illustrates the consequences of these values:
         The transfer of a single photon between two adjacent stations placed $\u{80}{km}$ apart takes $\u{0.4}{ms}$, using the speed of light in fiber, $c = \uf{2 \cdot 10^5}{km}{s}$, and has a transmission probability of~$\u{2.5}{\%}$.
         A scheme in which the successful or failed transmission is signaled back and only then triggers the next action will take $\u{32}{ms}$ on average---completely neglecting the time for any local operations, further meta\hyp communication, and the practical difficulties that come with quantum memories.

         In classical communication, these values do not restrict the actual bandwidth.
         The intensity of a classical signal is high enough that a reduction by$\u{98}{\%}$ still amounts to a successful transmission.
         Hence, there is no need to signal back whether a transmission was actually successful; instead, the next photon packet can be sent almost immediately after its predecessor.
         This is why the distance between two classical communication parties only has an impact on \emph{latency}, i.e., initialization of the communication, but not on the \emph{rate} of information transfer.

      \subsection{Error correction}
         The same thought carries over to the quantum regime: if the successful transmission of a signal needs to be signaled back to the previous station before it can carry on with the next signal, the rate is severely bound by the distance between those stations.
         Hence, a truly scalable QKD system must be able to avoid two\hyp way communication between repeater stations under all circumstances; or to make it non\hyp blocking using an enormous number of quantum memories, which, despite recent achievements~\cite{bib:Bradley2019}, is a huge challenge.
         Note this argument does not forbid the receiver from sending some required information back to the sender when the message arrives, as long as there is only this one round of backwards communication affecting only classical information.
         This means that the sender must only store all the classical information until the confirmation or failure signal from the receiver arrives.
         Hence, the availability of large (compared to the quantum case, quasi\hyp infinite) classical memory allows us to count this waiting time of the sender as an initial latency with no impact on the rate: there is no need to wait for the heralding signal from the receiver before sending the next state.
         This is based on the assumption that signals can be sent arbitrarily close to each other, which is an idealization; but still, the possibility to adjust the waiting time between signals is now a way to combat negative impacts of failed transmissions on the rate, which is not possible if the successful arrival has to be heralded between each station.

         \subsubsection{Traditional quantum error correcting codes}\label{sec:qecc}
            Whereas quantum repeaters of the first and second generation rely on backwards communication, third\hyp generation repeaters employ quantum error correcting codes (QECCs)~\cite{bib:Shor1995,bib:Steane1996,bib:Knill1997,bib:Gottesman1997} to circumvent attenuation without the requirement of additional feedback communication~\cite{bib:Muralidharan2016}.
            An $\code{n, k, d}$ QECC is a $2^k$\hyp dimensional subspace of a $2^n$\hyp dimensional Hilbert space; or, it stores the content of $k$~\emph{logical} qubits into $n$~\emph{physical} qubits.
            It is able to correct against $\bigl\lfloor\frac{d -1}{2}\bigr\rfloor$ arbitrary single\hyp qubit errors, among which we can also count deletions, i.e., losses whose position is unknown.
            Erasures---losses with known position---can be corrected for on up to $d -1$ qubits~\cite{bib:Grassl1997}.
            Hence, the idea behind employing an error correcting code against attenuation effects is that it is sufficient to successfully transfer a lower number of bits instead of the full number, $n$, while still completing the action successfully.

            In this paper, we will not be concerned with errors due to faulty components or dark counts---we will only consider the errors introduced by channel attenuation.
            Our main focus will be on a ``reduced'' erasure scenario: While we will assume that lost qubits are known not to have arrived, once all those arrived qubits are bundled together, we will not exploit any further knowledge of which particular combination of qubits the bundle pertained to.
            We will explain and justify this model more in detail in \cref{sec:reder}.
            We assume that the loss of every photon is independent from all the others, and they all occur with the same probability.

            Since QECCs are required to deterministically restore the full state, the no\hyp cloning theorem~\cite{bib:Park1970,bib:Wootters1982,bib:Dieks1982} limits them to tolerate only error rates below $\u{50}{\%}$, or, in our case, a QECC requires $p_{\mathrm{trans}} > \frac12$.
            Translated into distance, neighboring stations must not be more than $\u{15}{km}$ apart.

            We note that while we used the $\code{n, k, d}$ notation commonly employed for stabilizer\hyp based (additive) quantum codes, there have been a few reports about better performance of some exemplary nonadditive codes~\cite{bib:Rains1997,bib:Smolin2007,bib:Yu2008}.
            However, apart from the fact that so far, only very low code distances were reported, they are still QECCs that are bound by the no\hyp cloning theorem.

         \subsubsection{Redundant parity encoding}\label{sec:rpe}
            An $\code{n, k, d}$ code is designed to be robust against errors, i.e., to allow for the correction of arbitrary quantum errors.
            For the particular issue of losses, other codes have been developed~\cite{bib:Grassl1997}.
            In particular, for use with quantum repeaters, a redundant parity encoding~\cite{bib:Ralph2005,bib:Munro2012} has been proposed.
            The general idea is to encode the state $\alpha \ket0 + \beta \ket1$ in two layers of encoding, giving (unnormalized)
            \begin{equation}
               \alpha \bigl[ \ket0^{\otimes m} + \ket1^{\otimes m} \bigr]^{\otimes n} +
               \beta \bigl[ \ket0^{\otimes m} - \ket1^{\otimes m} \bigr]^{\otimes n}\text.
            \end{equation}
            At the lower level, we have a block consisting of $m$ qubits prepared in a GHZ state~\cite{bib:Greenberger1989,bib:Bouwmeester1999}; at the higher level, there are $n$ of those blocks.
            Now note that measuring an arbitrary qubit in the computational basis effectively disentangles its whole embedding $m$-qubit block from all the other blocks.
            Therefore, provided at least one photon from each block survives, all blocks can be disentangled from each other by measuring one photon from each but one block in the computational basis.
            From the block that was left unchanged, we require to have access to all its $m$ qubits.
            It is now in the (unnormalized) state
            \begin{equation}
               \alpha \bigl[ \ket0^{\otimes m} + \ket1^{\otimes m} \bigr]
               \pm \beta \bigl[ \ket0^{\otimes m} - \ket1^{\otimes m} \bigr]\text,
            \end{equation}
            where the sign depends on the sum of all previous measurements and can therefore be corrected.
            We end up with the logical state of the qubit, which may be decoded or processed further.

            For some values of $m$ and $n$ this actually describes a QECC---for instance, the Shor code~\cite{bib:Shor1995} is given by $m = n = 3$---while for others, it does not, the most simple example being $n = 1$, $m = 2$.

            Unfortunately, the previous analysis in \cite{bib:Munro2012} has neglected that the two requirements are competitive (see \cref{app:rpe} for details): large~$m$ and small~$n$ are required to have a successful disentangling operation, while large~$n$ and small~$m$ are needed to then successfully restore the initial state from the disentangled one.
            The total success probability is given by
            \begin{equation}
               p\Bigl( \substack{\geq 1\text{ physical qubits} \\
                                 \text{in each block}} \Bigr) -
               p\Bigl( {\underbrace{
                           \substack{1\text{ to }(m -1)\text{ physical qubits} \\
                                     \text{in each block}}
                         }_{\substack{=\text{ never a complete block} \\
                                      \text{among the left set}}}
                        } \Bigr)\text,
            \end{equation}
            which, defining $\bar p_{\mathrm{trans}} \coloneqq 1 - p_{\mathrm{trans}}$, is
            \begin{equation}
               \bigl(1 - \bar p_{\mathrm{trans}}^m\bigr)^n -
               \bigl(1 - \bar p_{\mathrm{trans}}^m - p_{\mathrm{trans}}^m\bigr)^n\text.
               \label{eqn:rpecorrect}
            \end{equation}

            We numerically scanned $(m, n) \in [1, 1000]^2 \intersect \Z^2$ and found that redundant quantum parity codes can only deliver higher transmission probabilities than direct transmission for $p_{\mathrm{trans}} \geq \u{54}{\%}$.

         \subsubsection{Cluster states}
            The use of cluster states was suggested to tolerate high error rates~\cite{bib:Varnava2006,bib:Varnava2007}.
            However, as these were mostly based on QECCs, the limit of $\u{50}{\%}$ failure also applies here.

            Recently, it has been pointed out that there are cluster states for which no foliation to a traditional stabilizer code exists~\cite{bib:Nickerson2018}.
            The existence of a cluster state with a $\u{55}{\%}$ erasure threshold was reported and it was pointed out that there is no fundamental limit.
            Although we do not use cluster state methods in this paper, our approach later will be motivated by this fact.

      \subsection{Beyond traditional error correction}\label{sec:multdof}
         The previous approaches share two common traits: within their scope of applicability, they correct deterministically; and they are based on qubits.

         Addressing the former, we note that the secret key rate in \cref{eqn:qkdrate} scales linearly with a probabilistic factor and has an entropic dependency on a quality factor.
         QECCs are \emph{deterministic} codes: their probabilistic element is purely based on the question---which is typically considered external to the code---of whether few enough errors happened, and in this case they are always able to deliver perfect quality.
         If too many errors occur, the code\hyp specific behavior is usually uncharacterized.
         We will ask the question of how much loss can be tolerated in a state if we allow the correction operation to only probabilistically succeed.
         Here, by ``success'' we do not require a perfect recovery, as \cref{eqn:qkdrate} has a non\hyp zero rate even if the states are slightly erroneous.
         Instead, we will investigate how to achieve an optimal combination of success probability and recovery quality with regard to the secret key rate.
         This is justified by the fact that we may signal back the success of the total operation to the sender without negatively impacting the rate.
         We note that a very recent study of probabilistic transformations from the point of view of resource theories also investigates the potential benefits of nondeterministic operations~\cite{bib:Kondra2021}.

         More out of tradition and its close relationship with classical error correction, QECCs are defined in terms of qubits.
         In \cite{bib:Munro2012}, it was noted that a photon has multiple degrees of freedom that can be exploited for the encoding.
         While using more degrees of freedom per photon implies that the photons that arrive will be able to carry more information, at the same time, the photons that do \emph{not} arrive will also \emph{lose} more information.
         For this reason, it is not immediately obvious if and in what manner going to higher dimensions will prove beneficial.
         We will investigate this possibility in \cref{sec:qudit} by moving from qubits to qudits; and indeed, for some configurations, we can report a significant improvement in the transmitted fidelity.

      \subsection{The ``reduced'' erasure model}\label{sec:reder}
         We previously introduced the ``reduced'' erasure model, which is the subject of study in this paper.
         It arises from the conventional, ``full'' erasure by dropping the information about which particular loss configuration occurred, while still removing all lost slots from the input to the correction circuit.
         We will illustrate this with an example.

         Imagine that we send three photons $1$, $2$, and~$3$; the full erasure information now consists in knowing which one of
         \begin{equation*}
            \bigl\{ \emptyset, \{ 1 \}, \{ 2 \}, \{ 3 \}, \{ 1, 2 \}, \{ 1, 3 \}, \{ 2, 3 \}, \{ 1, 2, 3 \} \bigr\}
         \end{equation*}
         arrived.
         In principle, the receiver may implement eight (or seven, disregarding the case of complete failure) different maps corresponding to all the different configurations.
         We will mostly only study the case with a given number of received photons (e.g., corresponding to the most likely event; but see \cref{sec:varyr}), but even if the receiver knows that, say, two photons arrived, it can still lead to three different maps.
         The process becomes simpler, both experimentally and for our numerical study, if the receiver only has to implement a \emph{single} correction map that takes two input states: this corresponds to the ``reduced'' erasure information.
         Still, the receiver must route the two successful arrivals into the proper input slots for the map, but after this photonic switch, only one circuit needs to be run.

         While the theoretical model of quantum erasures is a very common one~\cite{bib:Bennett1997,bib:Grassl1997,bib:Gottesman1997}, recent experimental progress actually makes it feasible for the use in quantum communication.
         For example, non\hyp destructive photon detectors~\cite{bib:Niemietz2021,bib:Distante2021} can be used to directly obtain all the erasure information; in principle, boosted teleportation schemes~\cite{bib:Ewert2014} should provide means to detect photons nondestructively with arbitrarily high success probabilities using linear optics and resource state generators alone.
         Alternatively, the necessary data may be obtained by storing the arriving photons in quantum memories in a heralding way~\cite{bib:Brekenfeld2020,bib:Bhaskar2020,bib:Kalb2015,bib:Michelberger2014}, which may have benefits if the quantum information is to be manipulated in this form anyway.

         We note that in all these situations, a suitable protocol will use classical communication and synchronized clocks to first establish what the receiver \emph{should} expect.
         Then, the erasure information is obtained by comparing the output of the aforementioned detectors or heralding signals with the expectations.
         Since this actually reveals the \emph{full} erasure information, we also see potential for extending our method to cover this case.
         However, since this will render the symmetry reduction in \cref{sec:reddim} impossible, additional insight is required in order to scale the full erasure case to larger systems.
         We plan to address this in a forthcoming publication.

   \section{Optimizing transmission}\label{sec:mult}
      The TGW~\cite{bib:Takeoka2014} and the PLOB~\cite{bib:Pirandola2017} bounds require us to use quantum repeaters for a scalable QKD network.
      However, current research is dominated either by repeater distances of $\u{1\,\text{--}\,2}{km}$~\cite{bib:Muralidharan2014,bib:Lee2019,bib:Alsina2021}, which are far beyond what the current infrastructure has to offer, or they require inter\hyp repeater two\hyp way communication over long distances together with a large number of quantum memories~\cite{bib:Simon2007,bib:Scott2017,bib:Nickerson2014}.

      The most effective way to securely transmit information would be to find an optimal QKD protocol (preparation, transfer, and key processing) for a given repeater sequence.
      In a highly connected quantum network, then the best possible sequence is selected as the chosen pathway.
      This approach is highly impractical, and we discard it in favor of a ``plug\hyp and\hyp play'' solution that---while not necessarily being globally optimal---does not require a different protocol whenever the routing between the parties differs:
      We are interested only in finding the optimal protocol that allows to establish a Bell state $\ket{\Phi^+} \propto \ket{00} + \ket{11}$ between two adjacent neighbors.
      This segment is then suitable for use in any specific communication chain and thus provides a lower bound on what is achievable, while still being much more easy to implement than a global protocol.

      \emph{Outline}---In the following subsections, we will develop our method to numerically find such optimal protocols.
      In \cref{sec:multi}, we will first establish the context in which our approach will fit.
      The most general formulation applicable to the ``reduced'' erasure scenario, previously introduced in \cref{sec:qecc,sec:reder}, will be given in \cref{sec:reduced}.
      Subsequently, in \cref{sec:reddim}, we analyze the setting in more detail: We will exploit a permutation symmetry in order to reduce the effective dimensionality of our systems and will end up with a scaling that is linear in the number of sent and received particles.

      While this greatly simplifies the optimization task, it is still a nonconvex problem.
      We will discuss two approaches that are suitable to solve this problem: nested optimization with convex subproblems (\cref{sec:consub}) and a method that is called \emph{convex iteration} (\cref{sec:conit}).
      In \cref{sec:pract}, we bring those approaches together: we describe advantages and disadvantages and detail an optimization pipeline that uses both methods, and which we found to give satisfying results with acceptable resource consumption.

      We will then discuss some choices that we made in our model and how to generalize them, from the assumption of sending qubits (\cref{sec:qudit}), the fixed value~$r$ of arriving particles (\cref{sec:varyr}), to our decision to only consider the ``reduced'' erasure setting instead of the full erasure knowledge (\cref{sec:fullerasure}).

      A brief discussion in \cref{sec:process} of how the resulting numerical data---which will typically lead to extremely complicated states and maps---may be simplified concludes the discussion of our methods.

      \subsection{Multiplexing and entanglement distillation}\label{sec:multi}
         A direct transmission success probability of $\u{2.5}{\%}$ is too low to allow for successful large\hyp scale QKD.
         However, it is not too low for multiplexing techniques to provide substantial advantage.

         Multiplexing was already suggested multiple times in the literature~\cite{bib:Collins2007,bib:Munro2010} and to some extent also demonstrated experimentally~\cite{bib:Li2019}.
         Although it is mostly found for all\hyp photonic implementations~\cite{bib:Azuma2015}, this is no necessity.
         The basic idea is simple: If we require the successful transmission of one photon, but we send $s$~copies instead, the probability that the transmission failed---which now only means that \emph{no} photon arrived---is given by $1 - \bar p_{\mathrm{trans}}^s$.
         By tuning~$s$, the success probability can be brought arbitrarily close to~$1$ at the expense of sending more and more copies.
         In particular, using the transmission probability for $\u{80}{km}$, we already need to multiplex with $28$~photons in order to have at least $\u{50}{\%}$ success probability.

         The difference to the codes presented before is that now each photon is independent of all others---but, as a consequence, the associated half of each copy has to be stored in its individual quantum memory.
         Additionally, the successful arrival of a copy has to be signaled back so that then, entanglement swapping can be performed on the correct pair.
         Since this is in strong conflict with the previously stated goal to avoid backwards communication between repeaters at all costs, we instead want to perform multiplexing with $s$~qubits while the sender only stores a \emph{single} qubit, which need not be processed further based on some outcome at the receiver's site\footnote{%
            In fact, a single unitary operation conditioned on a measurement outcome at the receiver's site is still permissible---since this correction can be accounted for only at the very end, as is the case for a chain of multiple entanglement swappings.
            However, we find that allowing for such an operation does not appear to lead to any improvement.
         }.

         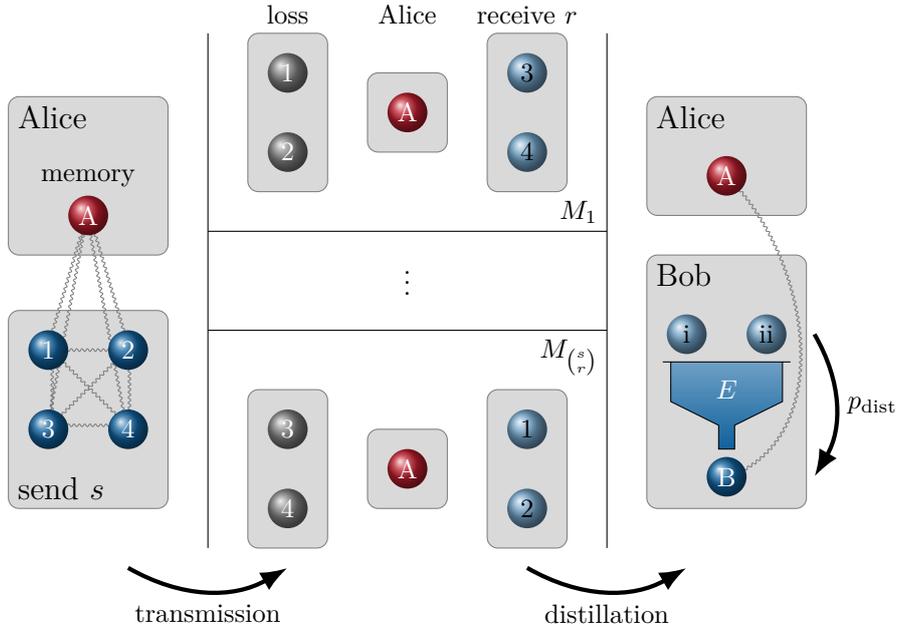
\begin{figure*}[t!]
            \centering
            \begin{tikzpicture}[scale=1.05]%[yscale=.8, every node/.append style={scale=.8}]
               \begin{scope}[shift={(0, 0)}]
                  \filldraw[draw=gray, fill=gray!30!white, rounded corners] (0, .2) rectangle (2, 2.2)
                     (0, -3) rectangle (2, -.5);
                  \node[anchor=north west, font=\large] at (0, 2.2) {Alice};
                  \node[anchor=south west, font=\large] at (0, -3) {send $s$};
                  \coordinate (A) at (1, .7);
                  \coordinate (S1) at (.5, -1);
                  \coordinate (S2) at (1.5, -1);
                  \coordinate (S3) at (.5, -2);
                  \coordinate (S4) at (1.5, -2);
                  \draw[gray, decoration={snake, amplitude=.5pt, segment length=2pt}, decorate]
                     (S1) -- (S2) -- (S3) -- (S1) -- (S4) -- (S3) (S4) -- (S2)
                     (A) -- (S1) (A) -- (S2) (A) -- (S3) (A) -- (S4);
                  \shade[shading=ball, ball color=alice] (A) circle[radius=.25] node[white] {$\mathrm A$} ++(0, .25) node[above] {memory};
                  \shade[shading=ball, ball color=bob] (S1) circle[radius=.25] node[white] {$1$};
                  \shade[shading=ball, ball color=bob] (S2) circle[radius=.25] node[white] {$2$};
                  \shade[shading=ball, ball color=bob] (S3) circle[radius=.25] node[white] {$3$};
                  \shade[shading=ball, ball color=bob] (S4) circle[radius=.25] node[white] {$4$};
               \end{scope}
               \draw (2.5, 3) -- (2.5, -3.5);
               \draw[ultra thick, -Latex] (1.5, -3.75) to[bend right=30] (3.5, -3.75);
               \node[anchor=north] at (2.5, -4.1) {transmission};
               \begin{scope}[shift={(3, 1)}]
                  \filldraw[draw=gray, fill=gray!30!white, rounded corners] (0, 0) rectangle (1, 2) (1.5, .5) rectangle (2.5, 1.5) (3, 0) rectangle (4, 2);
                  \node[anchor=south] at (2, 2) {Alice};
                  \node[anchor=south] at (.5, 2) {loss};
                  \node[anchor=south] at (3.5, 2) {receive $r$};
                  \coordinate (A) at (2, 1);
                  \coordinate (S1) at (.5, 1.5);
                  \coordinate (S2) at (.5, .5);
                  \coordinate (S3) at (3.5, 1.5);
                  \coordinate (S4) at (3.5, .5);
                  \shade[shading=ball, ball color=alice] (A) circle[radius=.25] node[white] {$\mathrm A$};
                  \shade[shading=ball, ball color=gray] (S1) circle[radius=.25] node[white] {$1$};
                  \shade[shading=ball, ball color=gray] (S2) circle[radius=.25] node[white] {$2$};
                  \shade[shading=ball, ball color=bob!60!white] (S3) circle[radius=.25] node {$3$};
                  \shade[shading=ball, ball color=bob!60!white] (S4) circle[radius=.25] node {$4$};
                  \node[anchor=north east] at (4.5, 0) {$M_1$};
                  \draw (-.5, -.5) -- (4.5, -.5);
               \end{scope}
               \node at (5, -.05) {$\vdots$};
               \begin{scope}[shift={(3, -3.5)}]
                  \draw (-.5, 2.75) -- (4.5, 2.75);
                  \node[anchor=north east] at (4.5, 2.75) {$M_{\binom sr}$};
                  \filldraw[draw=gray, fill=gray!30!white, rounded corners] (0, 0) rectangle (1, 2) (1.5, .5) rectangle (2.5, 1.5) (3, 0) rectangle (4, 2);
                  \coordinate (A) at (2, 1);
                  \coordinate (S1) at (.5, 1.5);
                  \coordinate (S2) at (.5, .5);
                  \coordinate (S3) at (3.5, 1.5);
                  \coordinate (S4) at (3.5, .5);
                  \shade[shading=ball, ball color=alice] (A) circle[radius=.25] node[white] {$\mathrm A$};
                  \shade[shading=ball, ball color=gray] (S1) circle[radius=.25] node[white] {$3$};
                  \shade[shading=ball, ball color=gray] (S2) circle[radius=.25] node[white] {$4$};
                  \shade[shading=ball, ball color=bob!60!white] (S3) circle[radius=.25] node {$1$};
                  \shade[shading=ball, ball color=bob!60!white] (S4) circle[radius=.25] node {$2$};
               \end{scope}
               \draw (7.5, 3) -- (7.5, -3.5);
               \draw[ultra thick, -Latex] (6.5, -3.75) to[bend right=30] (8.5, -3.75);
               \node[anchor=north] at (7.5, -4.1) {distillation};
               \begin{scope}[shift={(8, .7)}]
                  \filldraw[draw=gray, fill=gray!30!white, rounded corners] (0, 0) rectangle (2, 1.5)
                     (0, -3.7) rectangle (2, -.5);
                  \node[anchor=north west, font=\large] at (0, 1.5) {Alice};
                  \node[anchor=north west, font=\large] at (0, -.5) {Bob};
                  \draw[gray, decoration={snake, amplitude=.5pt, segment length=2pt}, decorate]
                     (1, .5) to[out=-45, in=20] (1, -3.3);
                  \shade[shading=ball, ball color=alice] (1, .5) circle[radius=.25] node[white] {$\mathrm A$};
                  \shade[shading=ball, ball color=bob!60!white] (.5, -1.5) circle[radius=.25] node {i};
                  \shade[shading=ball, ball color=bob!60!white] (1.5, -1.5) circle[radius=.25] node {ii};
                  \shadedraw[shading=axial, top color=bob!60!white, bottom color=bob]
                     (.2, -1.85) -- ++(1.6, 0)
                     ++(-.1, 0) -- ++(0, -.5) -- ++(-.6, -.3) -- ++(0, -.3) -- ++(-.2, 0) --
                     ++(0, .3) -- ++(-.6, .3) -- ++(0, .5);
                  \node[white] at (1, -2.2) {$E$};
                  \shade[shading=ball, ball color=bob] (1, -3.3) circle[radius=.25] node[white] {$\mathrm B$};
                  \draw[ultra thick, -Latex] (2.1, -1.5) to[bend left=30] (2.1, -3.3);
                  \node[anchor=west] at (2.4, -2.4) {$p_{\mathrm{dist}}$};
               \end{scope}
            \end{tikzpicture}
            \caption{%
               Visualization of the setting.
               After preparation at Alice's site, $s$~qubits are sent though the channel, but only~$r$ arrive.
               All possible configurations, corresponding to the loss maps $\{ M_i \}_i$, are independent and equally likely.
               A local distillation map~$E$ then aims at producing a high\hyp quality Bell pair between Alice and Bob with probability $p_{\mathrm{dist}}$; this map does not know which particles were lost.
               We visualize the important initial and final correlations by wavy lines, but suppress them at the intermediate stages.
               While we name the individual parties ``Alice'' and ``Bob,'' they may also be repeaters.
            }
            \label{fig:process}
         \end{figure*}
         Put into the larger context, every repeater (and the sender) creates an $(s +1)$-qubit state~$\rho$, stores a single qubit and sends the remaining $s$~qubits to its neighboring node.
         This node, due to losses, will receive $r \leq s$~of them (typically, $s \gg 1$ while $r \sim 1$) and execute a local postprocessing map~$E$ to convert them to a single qubit that, together with the stored one at the previous site, has the highest possible overlap with a pure Bell state.
         These steps are visualized in \cref{fig:process}.
         Finally, the repeater performs entanglement swapping with \emph{its} stored single qubit and the one resulting from the postprocessing.
         Every operation has to produce states with sufficiently high fidelity and success probability (but, contrary to QECCs, not unit probability), so that errors are kept small and success needs not be signaled back to the previous station.
         Since the receiver is still allowed to signal back the successful arrival to the sender, even probabilities considerably different from one are permissible, at the expense of more classical memory.

         Phrased in this way, the repeaters have to carry out an entanglement distillation procedure.
         When we later discuss the possibility of sending higher\hyp dimensional states, it will become clear that even the case $r = 1$---which has local filtering operations~\cite{bib:Verstraete2001} as the optimal procedure for the case of qubits---should not \emph{a priori} be dismissed.

         There exist numerous entanglement distillation protocols, e.g., \cite{bib:Bennett1996,bib:Deutsch1996,*bib:Deutsch1998,bib:Macchiavello1998,bib:Metwally2002,bib:Feng2000,*bib:Okrasa2008,*bib:Feng2008,bib:Metwally2006,bib:Feng2012,bib:Zhou2020,bib:Hu2021}, approaches to a theory of entanglement distillation~\cite{bib:Dehaene2003,bib:Bombin2005,bib:Hostens2006}; and upper bounds~\cite{bib:Rozpedek2018} as well as numerical tools to incorporate experimental constraints~\cite{bib:Krastanov2019,bib:Jansen2021} have been reported recently.
         However, finding the optimal protocol with respect to \emph{both} the prepared state and the distillation map, but now with a given channel model and certain protocol restrictions (success probability, available erasure information, \dots) is an unsolved task; and indeed, for the particular issue we face here, it is not advantageous to use known protocols.
         They assume an input state which is mostly unknown, apart from a single parameter such as the fidelity or some entanglement measure.
         Here, we are free to prepare an arbitrary state at the sender's site; and given the particular case of channel loss, we then also know precisely which state the receiver will get.
         We will compare our results with known distillation protocols in \cref{sec:compare}.

      \subsection{Optimizing the ``reduced'' erasure setting}\label{sec:reduced}
         We will formulate our problem in a way that captures most of what is possible in the ``reduced'' erasure setting previously described; but we will also explore some possible extensions later on.

         Let $\{ M_i \}_i$ be a set of completely positive~(CP) and trace preserving maps, where each $M_i$ corresponds to a different combination of losing $s - r$ out of $s$ subsystems; mathematically, they are given by the partial trace maps on the lost subsystems.
         Let $p_{\mathrm{dist}} \in (0, 1]$ be an external parameter that denotes the desired distillation success probability, and let $\ket{\Phi^+} \propto \ket{00} + \ket{11}$ be a Bell state.
         Finally, let $C(E)$ denote the Choi state of the CP trace\hyp nonincreasing distillation map $E$.
         The task is then to solve
         \begin{equation}
            \left\{\begin{aligned}
               & \centerasif{=}{\max_{E, \rho} F_{\mathrm{min}}} \\
               & \centerasif{=}{\text{subject to}} \\
               F_{\mathrm{min}}
               & \leq \frac{1}{p_{\mathrm{dist}}} \braket{\Phi^+ | (
                  \openone_{\mathrm A} \otimes E
               ) \compose M_i [\rho] | \Phi^+} \ \forall i \\
               p_{\mathrm{dist}}
               & = \tr\bigl( (
                      \openone_{\mathrm A} \otimes E
                   ) \compose M_i [\rho] \bigr) \ \forall i \\
               \rho
               & \succeq 0, \quad
               \tr\rho
                 = 1 \\
               C(E)
               & \succeq 0, \quad
               \tr_{\mathrm B} \compose E
                  \preceq \openone_{\mathrm R}\text.
            \end{aligned}\right. \label[program]{eqn:blp}
         \end{equation}
         Here, $\openone_S$~is the identity map acting on a system~$S$, $\compose$~denotes composition, and the partial order with respect to the positive semidefinite cone is indicated by $\succeq$ and~$\preceq$.
         We will traditionally label the two parties Alice and Bob, although in most cases, they will be repeater nodes.
         Following this convention, the state $\rho$ is defined on the systems~$\mathrm A$ and~$\mathrm S$; each $M_i$ maps $\mathrm S$ to $\mathrm R$\footnote{%
            In fact, depending on which combination of subsystems arrived, we get different $\mathrm R_i$; but in the ``reduced'' erasure scenario described in \cref{sec:qecc,sec:reder}, we identify all those different $\mathrm R_i$ with each other.%
         }; and finally, the map $E$ turns $\mathrm R$ into $\mathrm B$.

         \Cref{eqn:blp} is a generalized bilinear and semidefinite program---which can also be written as an indefinite quadratically constrained and semidefinite program---obtaining a global solution is NP\hyp hard.

         Note that \cref{eqn:blp} is different from the standard formulation of entanglement distillation, in that it allows one party to perform global operations \emph{before} sending the state, which is encoded in the state preparation procedure.
         Traditional distillation protocols instead assume that the two parties are given parts of a rather arbitrary state and have to make the best out of it; however, if the distribution state is actually under control, not exploiting this power relinquishes a potential advantage.

         In the following subsections, we will detail how to reduce the dimensionality of \cref{eqn:blp} and show how to approach the problem numerically in an efficient way.
         We then proceed to give details on the actual implementation and results.
         \clearpage

         \emph{Alternatives.}
         We note that while the formulation of \cref{eqn:blp} is natural and the favorable reduction in dimensionality appears to be a direct consequence, it is not the \emph{only} natural formulation.
         We might demand not the \emph{minimum}, but the \emph{average} fidelity to be maximized.
         Instead of fixing each \emph{individual} success probability, we might choose to fix only the \emph{overall} success probability.
         While these changes to the problem formulation are certainly well\hyp justified, they do not enable a reduction in dimensionality, so that we will not consider them.
         Finally, we also did not allow for the conditional execution of a single unitary Alice's site, depending on a measurement outcome at Bob's, as we did not find any improvement using this freedom.

      \subsection{Reducing dimensionality}\label{sec:reddim}
         Even if \cref{eqn:blp} were convex, note that \smash{$\rho \in \mathcal B(\C^{2^{s +1}})$} and \smash{$C(E) \in \mathcal B(\C^{2^{r +1}})$}, which means that the program size scales exponentially in the number of sent or received particles.
         Solving the problem for more than a handful of photons would be impossible.

         However, note that we optimize the minimum of the fidelity over all possible losses, and every loss channel has to give rise to the same probability.
         Consequently, the optimal $\rho$ and $C(E)$ will be graded in terms of the worst loss configuration.
         We can thus make the following \emph{ansatz}:
         \begin{gather}
            \rho
            = \sum_{a_1, a_2 = 0}^1 \sum_{i, j = 0}^s
                 \rho_{a_1, a_2}(i, j) \ketbra{a_1 D^s_i}{a_2 D^s_j}\text,
            \label{eqn:symstate}
            \shortintertext{where}
            \ket{D^s_i}
            \coloneqq \frac{1}{\sqrt{\binom si}} \sum_{\substack{x_1, \dotsc, x_s \in \{ 0, 1 \} \\ \sum_k x_k = i}} \ket{x_1, \dotsc, x_s}
         \end{gather}
         is a Dicke state~\cite{bib:Bergmann2013}---i.e., we assume that the coefficients in the $2^{s +1} \times 2^{s +1}$ matrix $\rho$ only depend on the stored qubit at Alice and the \emph{number} of set bits that arrive at Bob, but not on their particular order.
         The structure is now similar to a permutation invariant quantum code~\cite{bib:Ouyang2014,bib:Ouyang2017}, without imposing the Knill\hyp Laflamme conditions~\cite{bib:Knill1997}.
         In this way, we exploit that favoring one particular loss configuration at the expense of another will actually \emph{reduce} the final (minimum) fidelity---so the best way to start is to make them all equal.
         While this reasoning is intuitive and also supported by numerical studies for small~$s$ and~$r$, a formal proof of optimality faces various problems; e.g., even if all reduced states are equal, they do not necessarily imply an original Dicke state; and there is no direct relation between the fidelities arising from an arbitrary state and its closest symmetrized version.

         The Dicke states are all orthonormal; we can thus reduce our attention to the $2(s +1) \times 2(s +1)$\hyp dimensional subspace that hosts our matrix $\rho$.

         In order to be able to work with these much smaller states, we must consider the action of any partial trace of $s - r$ qubits, excluding the one stored at Alice's site, on the full state.
         It is given by (in \cref{app:partracedit}, we explicitly derive a more general version)
         \begin{multline}
            \tr_{\mathrm{loss}} \rho
            = \sum_{a_1, a_2 = 0}^1 \sum_{k, \ell = 0}^r \sum_{i = 0}^{s - r}
               \binom{s - r}{i} \sqrt{\frac{\binom rk \binom r\ell}{\binom{s}{k + i} \binom{s}{\ell + i}}} \cdot {} \\
               \rho_{a_1, a_2}(k + i, \ell + i) \ketbra{a_1 D^r_k}{a_2 D^r_\ell}\text.
         \end{multline}

         This structure already shows that the exact same symmetry argument will be valid for the Choi state $C(E)$ of the distillation map, which will take the form
         \begin{equation}
            C(E)
            = \sum_{b_1, b_2 = 0}^1 \sum_{i, j = 0}^r c_{b_1, b_2}(i, j) \ketbra{b_1 D^r_i}{b_2 D^r_j}\text.
         \end{equation}

         \noindent We define the tensor
         \begin{equation}
            D_{k, \ell; k', \ell'}
            \coloneqq \delta_{k' - k, \ell' - \ell}
                 \binom{s - r}{k' - k}
                 \sqrt{\frac{\binom rk \binom r\ell}{\binom s{k'} \binom s{\ell'}}}
            \label{eqn:dtens}
         \end{equation}
         and find
         \begin{multline}
            E\bigl[\tr_{\mathrm{loss}}\rho\bigr]
              = \sum_{a_1, a_2 = 0}^1 \sum_{b_1, b_2 = 0}^1
                  \ketbra{a_1 b_1}{a_2 b_2} \cdot {} \\
                  \sum_{k, \ell = 0}^r \sum_{k', \ell' = 0}^s
                  c_{b_1, b_2}(k, \ell) D_{k, \ell; k', \ell'}
                  \rho_{a_1, a_2}(k', \ell') \text.
         \end{multline}

         The symmetry considerations have allowed us to reduce the number of (non\hyp slack) degrees of freedom in the optimization from $2^{2(s +1)} + 2^{2(r +1)}$ to $4(s +1)^2 + 4(r +1)^2$.
         An immediate side\hyp effect is that we no longer have to consider $\binom sr$ constraints on both fidelity and probability arising from all possible loss channels, as they automatically all give rise to the same final state.

         The problem now reads
         \begin{equation}
            \left\{\begin{aligned}
               \mathrlap{\max_{E, \rho} \frac{1}{p_{\mathrm{dist}}} \braket{\Phi^+ | E\bigl[\tr_{\mathrm{loss}}\rho] | \Phi^+}}\hphantom{\tr_{\mathrm{out}} E}
               & \\
               & \centerasif{=}{\text{subject to}} \\
               p_{\mathrm{dist}}
               & = \tr \compose E\bigl[\tr_{\mathrm{loss}}\rho \bigr] \\
               \rho
               & \succeq 0 \\
               \tr\rho
               & = 1 \\
               C(E)
               & \succeq 0 \\
               \tr_{\mathrm B} \compose E
               & \preceq \openone_{\mathrm R}\text.
            \end{aligned}\right.\qquad \label[program]{eqn:blpsmall}
         \end{equation}

         The program itself does not explicitly introduce complex numbers; but of course, $\rho$ and $C(E)$ are hermitian, not necessarily real.
         Based on the fact that at least for every stabilizer code, there is an equivalent real\hyp valued version~\cite{bib:Rains1999} and preliminary numerical studies with small parameters $s$ and $r$ that always lead to real\hyp values solutions, we will assume that both $\rho$ and $E$ are real\hyp valued, reducing the number of degrees of freedom to $(s +1) (2s +3) + (r +1) (2r +3)$.
         While dropping this requirement may open up possibilities for improvement on the results presented here, it considerably helps to keep the numerics tractable.
         Almost all numerical solvers work with real inputs only; hence hermitian matrices must be represented in a double\hyp sized real matrix with block\hyp symmetry constraints.
         Convex solvers typically require a time that is at least cubic in the size of the semidefinite matrices involved, which is a huge drawback; and for nonconvex solvers, the more variables are involved, the less confidence can be obtained about global optimality when some termination criteria are met.

      \subsection{Reformulation: convex subproblems}\label{sec:consub}
         We investigated various approaches to obtain good solutions for \cref{eqn:blpsmall}.
         Obviously, \cref{eqn:blpsmall} turns into a convex problem once either $E$ or $\rho$ are fixed.
         This leads to two possible reformulations:
         \begin{subequations}
            \begin{enumerate}
               \item given $E$, find $\rho$:
                  \begin{equation}\label[program]{eqn:noncon1}
                     \left\{\begin{aligned}
                        \mathrlap{\max_{\rho} \frac{1}{p_{\mathrm{dist}}} \braket{\Phi^+ | E\bigl[\tr_{\mathrm{loss}}\rho] | \Phi^+}}\hphantom{\tr_{\mathrm{out}} E}
                        & \\
                        & \centerasif{=}{\text{subject to}} \\
                        p_{\mathrm{dist}}
                        & = \tr \compose E\bigl[\tr_{\mathrm{loss}}\rho \bigr] \\
                        \rho
                        & \succeq 0 \\
                        \tr\rho
                        & = 1
                     \end{aligned}\right.\qquad
                  \end{equation}
               \item given $\rho$, find $E$:
                  \begin{equation}\label[program]{eqn:noncon2}
                     \left\{\begin{aligned}
                        \mathrlap{\max_{E} \frac{1}{p_{\mathrm{dist}}} \braket{\Phi^+ | E\bigl[\tr_{\mathrm{loss}}\rho] | \Phi^+}}\hphantom{\tr_{\mathrm{out}} E}
                        & \\
                        & \centerasif{=}{\text{subject to}} \\
                        p_{\mathrm{dist}}
                        & = \tr \compose E\bigl[\tr_{\mathrm{loss}}\rho \bigr] \\
                        C(E)
                        & \succeq 0 \\
                        \tr_{\mathrm B} \compose E
                        & \preceq \openone_{\mathrm R}
                     \end{aligned}\right.\qquad
                  \end{equation}
            \end{enumerate}
            \label[programs]{eqn:noncon}
         \end{subequations}

         Both reformulations can be solved in total by nonconvex, and in general local, optimization methods that, in each iteration, in turn call a convex solver for the convex subproblem.
         Note that for the nonconvex \cref{eqn:noncon2}, we can make use of another insight: for sure, the best initial state will be pure; mixed states can be interpreted as a loss of information of which pure state was prepared, which can never increase the final fidelity.
         This argument does not hold for $C(E)$, which in fact will turn out to be mixed at the optimum as soon as $r > 2$---the optimal measurement is not projective.
         Hence, while we have to optimize over about $(r +1)(2r +3)$ nonconvex degrees of freedom in \cref{eqn:noncon1}, we only need to consider about $2(s +1)$ nonconvex degrees of freedom in \cref{eqn:noncon2}.
         Of course, this is counteracted by the expected orders of magnitude $r \sim 1$, $s \gg 1$ of the parameters.
         Still, all nonconvex optimization algorithms we tried turned out to cope tremendously better with a state vector than with a general matrix optimization.

      \subsection{Reformulation: convex iteration}\label{sec:conit}
         The nonconvexity in \cref{eqn:blp} stems from the fact that $E\bigl[\tr_{\mathrm{loss}}\rho\bigr]$ contains products of $\rho$ and $E$.
         Note
         \begin{equation}
            x = y z
            \Leftarrow \rk \mat{a & x & y \\
                                x & b & z \\
                                y & z & 1} = 1\text,
         \end{equation}
         i.e., we can rewrite any product as a rank\hyp one constraint, which is still non\hyp convex.

         Relaxing the rank\hyp one by a positive semidefinite constraint removes any correlation between $x$ and $y z$: for all $x$, $y$, $z$, there exist $a$, $b$ such that the matrix is positive semidefinite.
         Therefore, the optimum of the relaxed problem will have nothing to do with the original problem.

         However, note that a feasibility problem that is, apart from rank(\hyp one) constraints, convex, can be re\hyp cast in the form of a sequence of convex problems which are non\hyp increasing in the sum of all but the largest singular value~\cite{bib:Dattorro2019,bib:Dedeoglu2016}.
         This sequence will, in general, not converge to zero, but we will later describe heuristics that allow to steer out of local rank minima.

         We will briefly outline the general procedure, closely following \cite[chapter~4.5]{bib:Dattorro2019}.
         First, \cref{eqn:blp} has to be cast into a feasibility problem, which we found to be superior to its alternative, multi\hyp objective optimization.
         Therefore, instead of maximizing the output fidelity, we fix a certain value for the fidelity and use the feasibility problem to check whether this is valid.
         The line $F \in [0, 1]$ can then be scanned to re\hyp obtain the maximization.
         Here, a binary search is not necessarily the most efficient way.
         In fact, a linear search can greatly outperform the binary search, as the convex iteration will profit from a good initial vector from the last round.
         Proceeding in small steps of, say, $\Delta F = 0.01$ will allow to quickly progress for a large range of fidelities with less than five iterations per check, whereas large steps could easily take hundreds of iterations.
         On top of this, detection of infeasibility is not a definite conclusion: the iteration might just be stuck in a local minimum.

         Now, our problem is to
         \begin{equation}
            \hspace{-2cm}\left\{\begin{aligned}
               & \centerasif{=}{\operatorname*{find}_{E, \rho} \, \{ G_i \}_i} \\
               & \centerasif{=}{\text{subject to}} \\
               F
               & = \frac{1}{p_{\mathrm{dist}}} \sum_i \mathcal F_i G_i^{1, 2} \tikzmark{usual1} \\
               p_{\mathrm{dist}}
               & = \sum_i \mathcal P_i G_i^{1, 2} \\
               \rho
               & \succeq 0, \quad
               \tr\rho
                 = 1 \\
               C(E)
               & \succeq 0, \quad
               \tr_{\mathrm B} \compose E
                 \preceq \openone_{\mathrm R} \tikzmark{usual2} \\
               \rk G_i
               & = \rk\renewcommand{\arraystretch}{1.15}
                       \mat{G_i^{1, 1} & G_i^{1, 2} & C(E)_{\mathfrak c(i)} \\
                            G_i^{1, 2} & G_i^{2, 2} & \rho_{\mathfrak r(i)} \\
                            C(E)_{\mathfrak c(i)}
                                       & \rho_{\mathfrak r(i)}
                                                    & 1} = 1 \ \forall i\text.
                   \hspace{-2cm}
            \end{aligned}\right.
            \label[program]{eqn:rankcon}
            \tikz[remember picture] \draw[overlay, decoration=brace, decorate] let \p1=(pic cs:usual1), \p2=(pic cs:usual2) in (\x1+5mm, \y1+4mm) -- (\x1+5mm,\y2-.5mm) node[midway, anchor=west, align=left, xshift=1mm] {${} \eqqcolon \boxast$};%
         \end{equation}%
         Here, we construct $3 \times 3$\hyp matrices~$G_i$ for every product of a $C(E)$ with a $\rho$~entry that appears as a term in the fidelity or probability.
         The vectors $\mathcal F$~and~$\mathcal P$ contain the corresponding coefficients, and the functions $\mathfrak c$~and~$\mathfrak r$ map the index~$i$ to the index in the matrices $C(E)$ and~$\rho$, respectively.

         A step in the convex iteration is made up of the semidefinite program
         \begin{equation}
            \hspace{-2cm}\left\{\begin{aligned}
               \MoveEqLeft\min_{E, \rho, \{ G_i \}_i}
               \sum_i \braket{G_i, W_i} \\
               \MoveEqLeft\text{subject to} \\
               \mathrlap{\boxast\text{, see \cref{eqn:rankcon}}}\hphantom{G_i}
               & \\
               G_i
               & \succeq 0 \ \forall i\text.
            \end{aligned}\right.
            \label[program]{eqn:cvxit}
         \end{equation}
         Here, $\braket{A, B} \coloneqq \tr(A^\top B)$ is the Hilbert\hyp Schmidt scalar product.
         Initially, we choose $W_i$ filled with~$1$ in every entry (but other choices, such as the identity, are also possible\footnote{%
            We found that the identity will converge faster; however, it is also more likely to end up with a non\hyp fixable stall than when starting with the matrix of~$1$\hyp entries.%
         }).
         Having found the optimal $\{ G_i \}_i$ for \cref{eqn:cvxit}, we decompose $G_i = U_i \operatorname{diag}(\lambda_i^1, \lambda_i^2, \lambda_i^3) U_i^\top$, where the columns of $U_i$ are made up of the eigenvectors of $G_i$, and $0 \leq \lambda_i^1 \leq \lambda_i^2 \leq \lambda_i^3$ are the corresponding eigenvalues.
         We obtain the direction vector for the next iteration by $W_i \mapsto U_i^\star U_i^{\star\top}$, where $U_i^\star$ has the last column dropped (i.e., the one corresponding to the largest eigenvalue).
         In this way, the quantifier of rank violation, $\sum_i \sum_{j = 1}^2 \lambda_i^j$, will never increase from iteration to iteration~\cite{bib:Dattorro2019}.
         However, it might not converge to zero.

         If the iteration stalls, i.e., the quantifier of rank violation does not decrease notably, we empirically find that the following perturbation allows to steer out of the local minimum; it slightly differs from the one proposed in \cite{bib:Dattorro2019}.
         Let $\vv r_i \in [0, 0.01]^2$ be a random vector and $\vv u_i^3$ the eigenvector of $U_i$ corresponding to the largest eigenvalue $\lambda_i^3$.
         Then, the next $W_i$ is given by $W_i = U^\star \bigl( \vv r_i^{\vphantom3} \vv u_i^{3\top} + U^{\star\top} \bigr)$.
         Note that the range of the random values is carefully chosen: larger ranges will completely deteriorate the iteration without any hope of recovery, while smaller ranges will not produce the desired liberation from the current valley.

         Also note that while the iteration was formulated in primal form, the corresponding dual form is actually more economic.

         \Cref{eqn:rankcon} is not the most efficient formulation of the problem in terms of rank constraints.
         Given that these constraints are nonconvex, we want to keep their number as small as possible.
         Observe that both $C(E)$ and $\rho$ are symmetric matrices, i.e., $c_{b_1, b_2}(k, \ell) = c_{b_2, b_1}(\ell, k)$ and likewise for $\rho$.
         To incorporate this symmetry, we will extend $D$ and, writing output indices before input indices, define
         \begin{widetext}
         \begin{equation}
            D_{A_1, B_1, A_2, B_2; b_1, k; b_2, \ell; a_1, k'; a_2, \ell'}
            \coloneqq \delta_{A_1, a_1} \delta_{B_1, b_1}
                \delta_{A_2, a_2} \delta_{B_2, b_2}
                D_{k, \ell; k', \ell'}
         \end{equation}
         in terms of the four\hyp index $D$\hyp tensor introduced in \cref{eqn:dtens}.

         We now have
         \begin{align}
            \braket{A_1 B_1 | E\bigl[\tr_{\mathrm{loss}}\rho\bigr] | A_2 B_2}
            & = \sum_{a_1, a_2 = 0}^1 \sum_{b_1, b_2 = 0}^1
                \sum_{k, \ell = 0}^r \sum_{k', \ell' = 0}^s
                   c_{b_1, b_2}(k, \ell)
                   D_{A_1, B_1, A_2, B_2; b_1, k; b_2, \ell; a_1, k'; a_2, \ell'}
                   \rho_{a_1, a_2}(k', \ell')\text.
            \intertext{Due to the symmetry of $C(E)$ and $\rho$, we can replace, leaving $E\bigl[\tr_{\mathrm{loss}} \rho\bigr]$ invariant,}
            D_{A_1, B_1, A_2, B_2; b_1, k; b_2, \ell; a_1, k'; a_2, \ell'}
            & \mapsto \frac14 \biggl(
                 \substack{\displaystyle
                    D_{A_1, B_1, A_2, B_2; b_1, k; b_2, \ell; a_1, k'; a_2, \ell'} +
                    D_{A_1, B_1, A_2, B_2; b_1, k; b_2, \ell; a_2, \ell'; a_1, k'} \hfill \\
                    \displaystyle{} +
                    D_{A_1, B_1, A_2, B_2; b_2, \ell; b_1, k; a_1, k'; a_2, \ell'} +
                    D_{A_1, B_1, A_2, B_2; b_2, \ell; b_1, k; a_2, \ell'; a_1, k'}
                 } \biggr)\text. \label{eqn:dsymm}
         \end{align}
         \end{widetext}
         Next, we can drop all entries from $D$ that refer to, say, strict lower triangles of the input and output matrices, and adjust the remaining terms appropriately.
         For example, we drop all entries from $D$ where $(a_1, k') \equiv a_1 (2s +1) + k' > a_2 (2s +1) + \ell' \equiv (a_2, \ell')$, and multiply by $2$ those with $(a_1, k') < (a_2, \ell')$.
         We then end up with a rank\hyp three tensor that maps the vectorized upper triangles of $C(E)$ and $\rho$ to the vectorized upper triangle of $E\bigl[\tr_{\mathrm{loss}} \rho\bigr]$.
         In fact, we can further reduce this by, in the output index, already summing over the elements required for the trace or overlap, respectively.
         The dimensions of the tensor are now $2$, $(r +1)(2r +3)$, and $(s +1)(2s +3)$.
         We perform singular value decompositions of the two matrices in the second and third indices; it turns out that it is possible to find joint left- and right\hyp singular bases to both of them.
         We get
         \begin{subequations}
            \begin{align}
               \tr E\bigl[\tr_{\mathrm{loss}}\rho\bigr]
               & = \sum_i d_i^{\mathrm{tr}}
                      \braket{c | d_i^c} \braket{d_i^\rho | \rho} \\
               \braket{\Phi^+ | E\bigl[\tr_{\mathrm{loss}}\rho\bigr] | \Phi^+}
               & = \sum_i d_i^\Phi
                      \braket{c | d_i^c} \braket{d_i^\rho | \rho}
            \end{align}
            \label{eqn:jointsvd}
         \end{subequations}
         All $(r +1)(2r +3)$ singular values $d_i^\Phi$ for the overlap are non\hyp zero; for the trace, we find $(r +1)(r +2)/2$ nonzero elements $d_i^{\mathrm{tr}}$.

         In this way, we only need to calculate the product of $(r +1)(2r +3)$ overlaps, which therefore also defines the number of rank constraints.
         In particular, it is independent of~$s$.

         Finally, note that we can reduce the number of iteration variables further.
         For this, we remark that
         \begin{equation}
            y z = \frac{(y + z)^2 - (y - z)^2}{4}\text;
         \end{equation}
         so any multiplicative constraint can instead be rewritten by two quadratic and some linear constraints.
         The quadratic constraints are of course also non\hyp convex; but the hypograph $t \geq x^2$ is a convex set.
         We therefore write
         \begin{align}
            \braket{c | d_i^c} \braket{d_i^\rho | \rho}
            & = \frac{\bigl(\braket{c | d_i^c} + \braket{d_i^\rho | \rho}\bigr)^2 - \bigl(\braket{c | d_i^c} - \braket{d_i^\rho | \rho}\bigr)^2}{4} \notag \\
            & \geq \frac{\bigl(\braket{c | d_i^c} + \braket{d_i^\rho | \rho}\bigr)^2 - t_i}{4}\text,
            \shortintertext{where}
            t_i
            & \geq \bigl(\braket{c | d_i^c} - \braket{d_i^\rho | \rho}\bigr)^2\text,
         \end{align}
         While this is of no use for the trace, where we need an exact equality---as the overlap with the Bell state is only a valid indicator for the fidelity if the trace is fixed---it is helpful for the overlap.
         Since we want to maximize the fidelity, we can alternatively use this lower bound for all singular values $d_i^\Phi$ where $d_i^{\mathrm{tr}} = 0$; and this requires only $2 \times 2$ rank matrices, reducing the number of iteration variables from six to three per matrix.

      \subsection{Practical optimization}\label{sec:pract}
         All reformulations have their individual advantages and disadvantages.
         Finding a suitable initial point for the nonconvex optimization is problematic, though it is of considerable help to express both $\rho$ and $C(E)$ in their singular bases found in \cref{eqn:jointsvd}.
         In order to reduce the likeliness of getting stuck in a local optimum, we would need to combine a local nonconvex optimization (such as Nelder\hyp Mead~\cite{bib:Nelder1965}) with global methods (such as basin\hyp hopping~\cite{bib:Wales1997,bib:SciPyBasinhopping}), which has dramatic impacts on the solving time of the nonconvex problems.
         The convex iteration, in turn, allows to quickly approach relatively good values, but in order to cross some ``fidelity barriers,'' hundreds or thousands of iterations are necessary until an appropriate random perturbation vector is found that allows to steer out of the local rank basin---even for relatively small system sizes.

         Here, we propose a combined approach: We first perform convex iteration until no significant rank improvements were possible for at least $50$~iterations.
         For large values of~$s$, we then use the output Choi state as an initial point for the nonconvex \cref{eqn:noncon1}, which does not work as well as \cref{eqn:noncon2}, but allows to steer into the right direction.
         To get a high\hyp quality solution, we finally always employ \cref{eqn:noncon2}, starting with the dominant eigenvector of~$\rho$ that resulted from the previous optimization.

         To improve reliability, the optimization proceeds independently on all probability sample points; after it has finished for probability $p_{\mathrm{dist}}$, the output state will be used as initial point for \cref{eqn:noncon2} with the neighboring probabilities $p_{\mathrm{dist}} \pm \Delta p_{\mathrm{dist}}$.
         If this improves the previously\hyp found solution by a certain threshold, we use the new result for the next probability in the same direction, until improvement is no longer achieved.

         As nonlinear optimization algorithm, we investigated all suitable methods implemented in the SciPy optimization library~\cite{bib:SciPyMinimize}, and also CMA~\cite{bib:PyCMA}.
         For \cref{eqn:noncon1}, only constrained methods (restricting all variables to $[-1, 1]$) were able to provide any useful result at all; and SLSQP~\cite{bib:Kraft1988} proved to be the superior one both in terms of speed and quality of the results by far.
         The most reliable algorithm for \cref{eqn:noncon2} turned out to be BFGS~\cite{bib:Broyden1970,bib:Fletcher1970,bib:Goldfarb1970,bib:Shanno1970,bib:Nocedal2006}.
         To perform the convex optimizations, we employ the MOSEK Optimization Suite~9.2.29~\cite{bib:Mosek}.
         We strongly recommend using an interior\hyp point solver for the convex iterations, as we experienced that iterations took longer and longer with a first\hyp order solver---in our case, SCS~\cite{bib:ODonoghue2016,bib:ODonoghue2019}---while naturally being much less accurate.

         All our implementations are available in a GitHub repository~\cite{bib:Desef2021}.
         While the individual classes starting with \texttt{Optimizer} correspond to the algorithms described in all parts of \cref{sec:mult} and can be used purely as an API, we provide two command line applications that may be used to generate all the numerical data backing our \cref{fig:fidsr,fig:fiddit}.
         We give a visual depiction of this optimization pipeline \cref{fig:python}.

         \begin{figure*}[t!]
            \begin{tikzalgorithm}[font=\footnotesize, tt/.style={font append=\ttfamily},
                                mainprog/.style={fill=yellow!20!white, draw=yellow},
                                subprog/.style={fill=cyan!20!white, draw=cyan},
                                code/.style={fill=gray!20!white, draw=gray}]
               \setlist{topsep=0pt, leftmargin=3.5mm}
               \algoblock[python Main.py <$d$> <$s$> <$r$>]{mainprog}{%
                  \algocommand{%
                     optional argument\hfill default\\[-3mm]
                     \begin{itemize}
                        \item \texttt{--outermap} (apply \cref{eqn:noncon1}?) \hfill \emph{unset}
                        \item \texttt{--erasure} (use ideas from \cref{sec:fullerasure}, not described in this figure) \hfill \emph{unset}
                        \item \texttt{--pmin=<value>} (minimum probability to scan) \hfill \texttt{0.01}
                        \item \texttt{--pmax=<value>} (maximum probability to scan) \hfill \texttt{1.00}
                        \item \texttt{--workers=<value>} (number of threads) \hfill machine-dependent
                     \end{itemize}
                  }%
                  \dimdef\colright{.49\linewidth}
                  \coordinate (savelastnode) at (lastnode.south west);
                  \algoblock{code}{
                     \algocommand[tt]{\textbf{for} $p \in \{ \text{pmin}, \text{pmin} + 0.01, \dotsc, \text{pmax} \}$:}
                     \algosub{%
                        \algoblock[OptimizerDickeConvexIteration.py]{subprog}{%
                           \algocommand{%
                              \begin{itemize}
                                 \item check: $F \in \{ 0.50, 0.51, \dotsc, 1 \}$ possible?
                                 \item at most 50 iterations per $F$
                                 \item algorithm: \cref{sec:conit}\\convex solver: MOSEK
                                 \item output: \texttt{bestState[p]}, \texttt{bestMap[p]}
                              \end{itemize}%
                           }
                        }
                        \algocommand[tt]{\textbf{if} outermap:}
                        \algosubblock[OptimizerDickeOuterMap.py]{subprog}{%
                           \algocommand{%
                              \begin{itemize}
                                 \item input: \texttt{bestMap[p]}
                                 \item algorithm: \cref{sec:consub,eqn:noncon1}\\outer solver: SciPy SLSQP; convex solver: MOSEK
                                 \item output: \texttt{bestState[p]}, \texttt{bestMap[p]}, \texttt{bestF[p]}
                              \end{itemize}
                           }
                        }
                        \algocommand[tt]{stateOptimize(p, bestState[p])}
                        \algocommand[tt]{\textbf{queue} stateOptimize(p - 0.01, bestState[p])}
                        \algocommand[tt]{\textbf{queue} stateOptimize(p + 0.01, bestState[p])}
                     }
                  }
                  \edef\colright{\linewidth}
                  \path (savelastnode) ++(.5\linewidth, 0) coordinate (lastnode);
                  \algoblock[stateOptimize(p, bestState)]{code}{
                     \algocommand[tt]{\textbf{if} p $\notin [\text{pmin}, \text{pmax}]$:}
                     \algosub{
                        \algocommand[tt]{\textbf{return}}
                     }
                     \algoblock[OptimizerDickeOuterState.py]{subprog}{
                        \algocommand{
                           \begin{itemize}
                              \item input: \texttt{bestState[p]}
                              \item algorithm: \cref{sec:consub,eqn:noncon2}\\outer solver: SciPy BFGS; convex solver: MOSEK
                              \item output: \texttt{newBestState}, \texttt{newBestMap}, \texttt{newF}
                           \end{itemize}
                        }
                     }
                     \algocommand[tt]{\textbf{if} newF $>$ bestF[p]:}
                     \algosub{
                        \algocommand[tt]{bestState[p] $\coloneqq$ newBestState}
                        \algocommand[tt]{bestMap[p] $\coloneqq$ newBestMap}
                        \algocommand[tt]{bestF[p] $\coloneqq$ newF}
                        \algocommand[tt]{\textbf{queue} stateOptimize(p-0.01, newBestState)}
                        \algocommand[tt]{\textbf{queue} stateOptimize(p+0.01, newBestState)}
                     }
                  }
               }
            \end{tikzalgorithm}
            \caption{%
               Algorithmic description of the optimization pipeline implemented in our GitHub repository~\cite{bib:Desef2021}.
               This depicts the main optimization program described in \cref{sec:pract}, and based on the algorithms discussed in \cref{sec:consub,sec:conit}; additionally, the parameter $d$ allows to control the dimension of the sent systems, as described in \cref{sec:qudit}.
               We also implemented a second optimization program \texttt{MainAllR.py}, which takes into account multiple values of~$r$ (see \cref{sec:varyr}); consult the command line help for its arguments.
               Note that both the main program with the \texttt{erasure} switch activated and the optimization over all values of~$r$ do not scale very well and should therefore only be considered for the case of qubits and small values of $s$ and~$r$.
            }
            \label{fig:python}
         \end{figure*}

      \subsection{Moving to qudits}\label{sec:qudit}
         In \cref{sec:multdof}, we already mentioned the possibility to exploit more photonic degrees of freedom by transmitting higher\hyp dimensional systems.
         The consequences are not immediately obvious: While now the arrival of a single photon can in principle transmit more information, also the loss of a single photon will lead to losing more information.
         Let $F_2(s, r, p)$ denote the fidelities obtained for qubits; then, assuming that a photon allows to implement $m \in \N$ qubits, the minimum achievable fidelity $F_{2^m}(s, r, p)$ is
         \begin{equation}
            F_{2^m}(s, r, p) \geq \max\bigl\{ F_2(s, r, p), F_2(m s, m r, p) \bigr\}\text,
         \end{equation}
         while the arrival probability is independent of $m$.
         Since this lower bound is obtained by either ignoring the new degrees of freedom or by considering them all independently, $F_{2^m}$ might actually be larger if we exploit the fact that the degrees of freedom pertaining to a single photon are always guaranteed to arrive together.

         \begingroup
            \parfillskip=0pt
            The formalism introduced before only needs to be adapted slightly in order to be able to deal with qudits.
            Closely following \cref{sec:reddim}, the dimensionality

         \endgroup

         \begin{widetext}
         \noindent reduction is now based on the \emph{ansatz}
         \begin{equation}
            \rho
            = \sum_{a_1, a_2 = 0}^1 \; \sum_{\vv i, \vv j \in [s]_d}
              \rho_{a_1, a_2}(\vv i, \vv j)
              \ketbra{a_1 D^s_{\vv i}}{a_2 D^s_{\vv j}}\text,
              \raisetag{5mm}
              \label{eqn:symstatedit}
         \end{equation}
         where
         \begin{equation}
            [s]_d \coloneqq \Biggl\{ \vv i \in \N_0^d : \sum_{m = 0}^{d -1} i_m = s \Biggr\}
            \label{eqn:defsd}
         \end{equation}
         and
         \begin{equation}
            \ket{D^s_{\vv i}}
            \coloneqq \frac{1}{\sqrt{\binom{s}{\vv i}}}
               \sum_{\substack{x_1, \dotsc, x_s \in \{ 0, \dotsc, d -1 \} \\ \lvert \{ j : x_j = m \} \rvert = i_m \ \forall m}} \ket{x_1, \dotsc, x_s}
            \label{eqn:defddicke}
         \end{equation}
         is the Dicke qudit state for $s = \abs{\vv i}$ particles in which the level $m$ occurs $i_m$~times ($m \in \{ 0, \dotsc, d -1 \}$).
         Here, we used the multinomial coefficient,
         \begin{equation}
            \binom{s}{\vv i}
            \equiv \binom{s}{i_0, \dotsc, i_{d -1}}
            \coloneqq \begin{dcases}
               \frac{s!}{i_0! \dotsm i_{d -1}!}
               & \substack{\displaystyle\text{if } \sum_{m = 0}^{d -1} i_m = s \\\displaystyle\hfill {} \land \vv i \succeq 0} \\\
               0
               & \text{else,}
            \end{dcases}
            \label{eqn:defmulti}
         \end{equation}
         where $\vv i \succeq 0$ is a component\hyp wise inequality.

         In \cref{app:partracedit}, we derive the action of the partial trace of $s - r$ qubits (any apart from the first one) on the full state \cref{eqn:symstatedit}.
         It is given by
         \begin{equation}
            \tr_{\mathrm{loss}} \rho
            = \sum_{\vv k, \vv\ell \in [r]_d}
              \sum_{\vv i \in [s - r]_d}
              \sum_{a_1, a_2 = 0}^1
                 \binom{s - r}{\vv i}
                 \sqrt{\frac{\binom{r}{\vv k} \binom{r}{\vv\ell}}
                            {\binom{s}{\vv k + \vv i} \binom{s}{\vv\ell + \vv i}}}
                 \rho_{a_1, a_2}(\vv k + \vv i,
                                 \vv\ell + \vv i)
                 \ketbra{a_1 D^r_{\vv k}}{a_2 D^r_{\vv\ell}}\text.
         \end{equation}
         With the help of the tensor
         \begin{align}
%               \braket{a_1 b_1 | E\bigl[\tr_{\mathrm{loss}}\rho\bigr] | a_2 b_2}
%               & = \sum_{\vv k, \vv\ell \in [r]_d}
%                   \sum_{\vv i \in [s - r]_d}
%                      \binom{s - r}{\vv i}
%                      \sqrt{\frac{\binom{r}{\vv k} \binom{r}{\vv\ell}}
%                                 {\binom{s}{\vv k + \vv i} \binom{s}{\vv\ell + \vv i}}}
%                      c_{b_1, b_2}(\vv k, \vv\ell)
%                      \rho_{a_1, a_2}(\vv k + \vv i, \vv\ell + \vv i)
%               \shortintertext{We now define}
         D_{\vv k, \vv\ell; \vv k', \vv \ell'}
%               & \coloneqq \sum_{\vv i \in [s - r]_d}
%                         \binom{s - r}{\vv i}
%                         \sqrt{\frac{\binom{r}{\vv k} \binom{r}{\vv\ell}}
%                                    {\binom{s}{\vv k + \vv i} \binom{s}{\vv\ell + \vv i}}}
%                         \delta_{\vv k', \vv k + \vv i}
%                         \delta_{\vv\ell', \vv\ell + \vv i} \\
            & \coloneqq \delta_{\vv k' - \vv k, \vv\ell' - \vv\ell}
                   \binom{s - r}{\vv k' - \vv k}
                   \sqrt{\frac{\binom{r}{\vv k} \binom{r}{\vv\ell}}
                              {\binom s{\vv k'} \binom s{\vv\ell'}}}\text,
         \end{align}
         we can now also express $E$---here, $C(E)$---in terms of Dicke states with coefficients $c_{b_1, b_2}(\vv k, \vv\ell)$ and finally find
         \begin{equation}
            E\bigl[\tr_{\mathrm{loss}}\rho\bigr]
              = \sum_{a_1, a_2 = 0}^1 \sum_{b_1, b_2 = 0}^1
                  \ketbra{a_1 b_1}{a_2 b_2}
                  \sum_{\vv k, \vv\ell \in [r]_d}
                  \sum_{\vv k', \vv\ell' \in [s]_d}
                  c_{b_1, b_2}(\vv k, \vv\ell) D_{\vv k, \vv\ell; \vv k', \vv\ell'}
                  \rho_{a_1, a_2}(\vv k', \vv\ell') \text.
         \end{equation}
         \end{widetext}

         While also here, the dimensionality reduction was able to greatly reduce the number of degrees of freedom, the situation is much more problematic than before.
         Since
         \begin{equation}
            \mbigl\lvert \big{[s]_d} \mbigr\rvert
            = \binom{s + d -1}{d -1}
              \simeq \frac{s^{d -1}}{(d -1)!}
         \end{equation}
         for large $s$, the complexity is now at least quadratic in $s$ and becomes intractable extremely quickly.
         Therefore, we can only optimize low numbers of $s$, $r$, and $d$ with the hope of finding optimal solutions.

      \subsection{Using more sophisticated processing}\label{sec:varyr}
         All previous schemes assumed that the number of received variables is some constant~$r$.
         However, due to the probabilistic nature of the transmission, the whole range $r \in \{ 0, \dotsc, s \}$ is in principle a valid choice\footnote{%
            Dark counts may in principle even allow for $r > s$, which we will not consider here.
            Note that every quantum signal may be accompanied by a classical heralding signal that allows to greatly diminish the effect of dark counts.%
         }.
         We might therefore decide to generalize our optimization problem to
         \begin{equation}
            \left\{\begin{aligned}
               \mathrlap{\max_{E_1, \dotsc, E_s, \rho} \frac{1}{p_{\mathrm{tot}}} \braket{\Phi^+ | \rho_{\mathrm f} | \Phi^+}}\hphantom{\tr_{\mathrm{out}} E}
               & \\
               & \centerasif{=}{\text{subject to}} \\
               \tr \rho_{\mathrm f}
               & = p_{\mathrm{tot}} \\
               \rho
               & \succeq 0 \\
               \tr\rho
               & = 1 \\
               C(E_j)
               & \succeq 0 \ \forall j \\
               \tr_{\mathrm B} \compose E_j
               & \preceq \openone_{\mathrm R} \ \forall j \\
               \rho_{\mathrm f}
               & = \sum_{j = 1}^s \binom sj p_{\mathrm{trans}}^j \bar p_{\mathrm{trans}}^{s - j} E_j\bigl[\tr_{\mathrm{keep}\,j}\rho \bigr]\text.
            \end{aligned}\right. \label[program]{eqn:blpmulrecv}
         \end{equation}

         Here, we replaced the probability of successful distillation~$p_{\mathrm{dist}}$ by the total success probability~$p_{\mathrm{tot}}$ that also incorporates the channel transmission.

         While we wrote ``for all~$j$'' in \cref{eqn:blpmulrecv}, for all practical purposes, we can fix a certain relative probability threshold; for arrival configurations with a lower probability than this threshold, we can disregard the additional maps.

         This approach is only moderately more time- and resource consuming when using \cref{eqn:noncon2}, but \cref{eqn:noncon1,eqn:cvxit} will suffer greatly under the increase of variables.
         On top of this comes the difficulty in experimentally implementing this scheme, where, depending on the number of arrived photons, a different map has to be applied.
         Finally, the procedure now explicitly contains the transmission probability---and since in real life, distances between repeater stations will vary, this may in fact imply different optimal maps and states depending on which repeater is the next in the chain.
         For these reasons, we will mostly refrain from using \cref{eqn:blpmulrecv}.
         When we briefly report some results from this optimization, we use the output states of the convex optimization routine for $r = 2$ and feed them as inputs to \cref{eqn:blpmulrecv}, using an outer nonconvex optimization over the initial state.

      \subsection{Exploiting full erasure knowledge}\label{sec:fullerasure}
         Our \emph{ansatz} crucially relied on the fact that we considered a ``reduced'' erasure scenario, i.e., we did not assume the information of which particular photon combination arrived to be available.
         This was reflected in the fact that in \cref{eqn:blp}, the distillation map~$E$ did not depend on the slot configuration index~$i$.
         We may choose to insert such a dependency, introducing a whole family of distillation maps~$E_i$ instead (or, combined with \cref{eqn:blpmulrecv}, even a two\hyp parameter family~$E_{j, i}$).
         In order to exploit this new information, the use of Dicke states now becomes prohibitive: due to their full symmetry with respect to a permutation of particles, we cannot improve our results in the setting of reduced dimensionality.
         Solving this problem for large values of~$s$ and~$r$ is not viable, but we are able to explore the landscape for small values mainly through \cref{eqn:noncon2}.

      \subsection{Processing the result data}\label{sec:process}
         The initial states~$\rho$ arising from the numerical optimizations will necessarily be pure.
         However, these $\rho = \ketbra\psi\psi$ still appear like arbitrarily complicated states.
         Note that the same final fidelity can be achieved if an arbitrary single\hyp qubit rotation with angle~$\alpha$ is carried out on the qubit stored at the sender's site and $s$~identical (transversal) single\hyp qubit rotations with an angle~$\beta$ (or their higher\hyp dimensional counterparts) are performed on the sent qubits.
         By minimizing the number of nonzero components in~$\ket\psi$ and the Choi state, this freedom in~$\alpha$ and~$\beta$ allows the complicated numerical output to still correspond to relatively tractable states.
         After applying this cardinality reduction to the full data, we re\hyp perform the optimization, removing the close\hyp to\hyp zero components from the beginning.
         This will speed up the optimization and lead to better results; however, in \cref{eqn:jointsvd}, the singular bases will no longer coincide for trace and overlap, so that we use the nested optimization approach.

         We note that the most promising way for improving both resource consumption and quality of the simulations would be to identify the relevant non\hyp zero coefficients beforehand.

   \section{Results}\label{sec:num}
      \subsection{Qubit scenario}
         We applied the optimization algorithms described in the previous sections to $1 \leq r \leq 10$.
         \Cref{fig:fidsr} shows the fidelities for various parameters $s$, $r$, and $p_{\mathrm{dist}}$ that can be achieved by explicit protocols using qubit transmission.
         The probabilities were sampled in $\u{1}{\%}$\hyp steps in the interval $[0.01, 1]$.

         The number of variables is small enough for these simulations that we can gain confidence in the actual optimality of the numerical lower bounds.
         The case $p_{\mathrm{dist}} = 1$ that is usually considered for QECC is now only the extreme point of lowest distillation fidelity; by lowering the distillation success probability, we get access to higher fidelities.
         As $p_{\mathrm{dist}}$ goes to zero, we may pass one or multiple points at which the fidelity curve is nonsmooth---the optimal initial state then takes a qualitatively different form that would have been suboptimal before.

         From \cite{bib:Grassl1997}, we know that there is a deterministic quantum erasure correcting code that is able to fully recover from a single erasure for $s = 4$, $r = 3$ (actually, this code falls into the same ``reduced'' class as we are studying), and that $s < 4$ does not allow for this.
         Our numerical data indeed confirms this.

         Note that for $s = 5$, $r = 3$, the best possible deterministic recovery gives a fidelity of $\u{80}{\%}$.
         Knowing that the perfect five qubit code~\cite{bib:Laflamme1996} has distance~$3$, it is actually possible to perfectly correct for two erasures deterministically; this gives an example of how we could improve if we assumed access to the full knowledge of which particular erasure occurred.
         While our numerical scheme hinted at in \cref{sec:fullerasure} is indeed able to provide such a code that gives a fidelity of $\u{99.87}{\%}$, this already stretches the capabilities of the algorithm; for $s = 6$, $r = 3$, there are so many variables that this advanced scheme delivers worse results than the reduced one.

         Using the technique described in \cref{sec:process}, in some cases---mainly for $p_{\mathrm{dist}} = 1$---we are able to derive analytical forms for the optimal states and maps based on the numerical data.
         We detail those results in \cref{app:analytic}.

      \subsection{Qudit scenario}
         When looking at the results for higher\hyp dimensional carriers such as qutrits, we cannot claim the same level of confidence in the optimality of the results because of the much higher number of degrees of freedom.
         Nevertheless, we compare some of the results in \cref{fig:fiddit}.

         We observe a very diverse behavior.
         Sometimes, moving to higher\hyp dimensional systems can pay off: Let us fix~$s$, then there is a minimal $r' \leq s$ that still allows to distill perfect pairs with unit probability for qubits (in the worst case, $r' = s$).
         The $(s, r' -1)$ combination will then typically profit greatly from moving to higher\hyp dimensional information carriers.
         Often, an advantage is visible only for distillation probabilities smaller than~$1$, but some combinations such as $s = 4$, $r = 2$ do not seem to have any potential for improvement (which has nothing to do with the $\u{50}{\%}$ loss mark---we find improvements for codes with less than, exactly, and more than $\u{50}{\%}$ loss).
         If qutrits are able to increase the fidelity, then ququarts can sometimes do even better; but sometimes, there is nothing to be gained.
         The question whether these issues pertain to the insufficient numerical means or whether we actually hit a barrier that cannot be overcome even by moving to still higher\hyp dimensional carriers remains open.

         Note that the full knowledge of which particular erasure happened, as mentioned in \cref{sec:fullerasure}, could in principle be encoded in some way in higher\hyp dimensional states.
         The new basis states might be used to flag the different components occurring in the Dicke state, which---depending on how well these flags can still be identified after the loss---could then be used to implement the different distillation maps $E_i$ as a single quantum map.
         However, this is \emph{not} what happens in the cases reported here: It would require the preparation to be made in the full basis, not the Dicke basis, i.e., there is no way to use the enlarged dimension in order to flag the different configurations in our scheme.

      \subsection{Comparison to known entanglement distillation schemes}\label{sec:compare}
         \afterpage{%
            % This is super crazy, but the only way the TeXLive version on the arXiv is able to output the figure instead of leaving a blank page.
            \addtocounter{figure}{-2}%
            \onecolumngrid
            \noindent\begin{minipage}{\linewidth}
               \begin{figure}[H]
                  \centering
                  \pgfplotsset{width=\axisdefaultwidth, height=6.6cm}
                  \pgfplotsforeachungrouped \s in {2, ..., 11} {%
                     \expandafter\edef\csname sty\s\endcsname{mpc\the\numexpr\s-2\relax}%
                  }%
                  \pgfplotsforeachungrouped \s in {20, 30, 40, 50} {%
                     \expandafter\edef\csname sty\s\endcsname{mpc\the\numexpr(\s-20)/10\relax, dashed}%
                  }%
                  \subfloat[$r = 1$]{%
                     \begin{tikzpicture}
                        \begin{axis}[xlabel=$p_{\mathrm{dist}}$, ylabel=$F$, legend pos=outer north east, enlarge x limits=false, axis on top, legend columns=2, transpose legend, legend to name=resleg, ymin=.46, ymax=1.04]
                           \addlegendimage{empty legend}
                           \addlegendentry{\raisebox{-2.6mm}{$s$\hspace{2mm}}}
                           \addlegendimage{empty legend}
                           \addlegendentry{}
                           \foreach \s in {2, ..., 11, 20, 30, 40, 50} {
                              \edef\cmd{%
                                 \noexpand\addplot[no marks, \csname sty\s\endcsname] table {./results/2-\s-1.dat};
                                 \noexpand\addlegendentry{$\s$}
                              }
                              \cmd%
                           }
                        \end{axis}
                     \end{tikzpicture}
                  }
                  \pgfplotsforeachungrouped \r in {2, ..., 6} {%
                     \ifodd\r\space\\[2mm]\else\hfill\fi%
                     \subfloat[$r = \r$]{%
                        \begin{tikzpicture}
                           \begin{axis}[xlabel=$p_{\mathrm{dist}}$, ylabel=$F$, enlarge x limits=false, axis on top, yticklabel pos/.expanded=\ifodd\r\space left\else right\fi, ymin=.46, ymax=1.04]
                              \edef\cmd{
                                 \noexpand\foreach \noexpand\s [count=\noexpand\i from \ifodd\r\space 2\else1\fi] in {\the\numexpr\r+1\relax, ..., 11, 20, 30, 40, 50} {
                                    \edef\noexpand\cmd{%
                                       \noexpand\noexpand\noexpand\addplot[no marks, \noexpand\csname sty\noexpand\s\noexpand\endcsname] table {./results/2-\noexpand\s-\r.dat};
                                    }
                                    \noexpand\cmd%
                                 }
                              }
                              \cmd
                           \end{axis}
                        \end{tikzpicture}
                     }
                  }\\[2mm]
                  \ref*{resleg}
                  \caption{%
                     Lower bounds on the maximally achievable fidelities for a pure loss channel with $s$ input qubits (given by the color coding) and $r$ output qubits, for various distillation success probabilities $p_{\mathrm{dist}}$.
                     Samples are taken at every integer percent value, but marks were omitted for clarity.
                     Every plot starts with $s = r +1$; if this line is not visible, it is obstructed by the $F = 1$ line of a higher $s$\hyp value.%
                  }
                  \label{fig:fidsr}
               \end{figure}
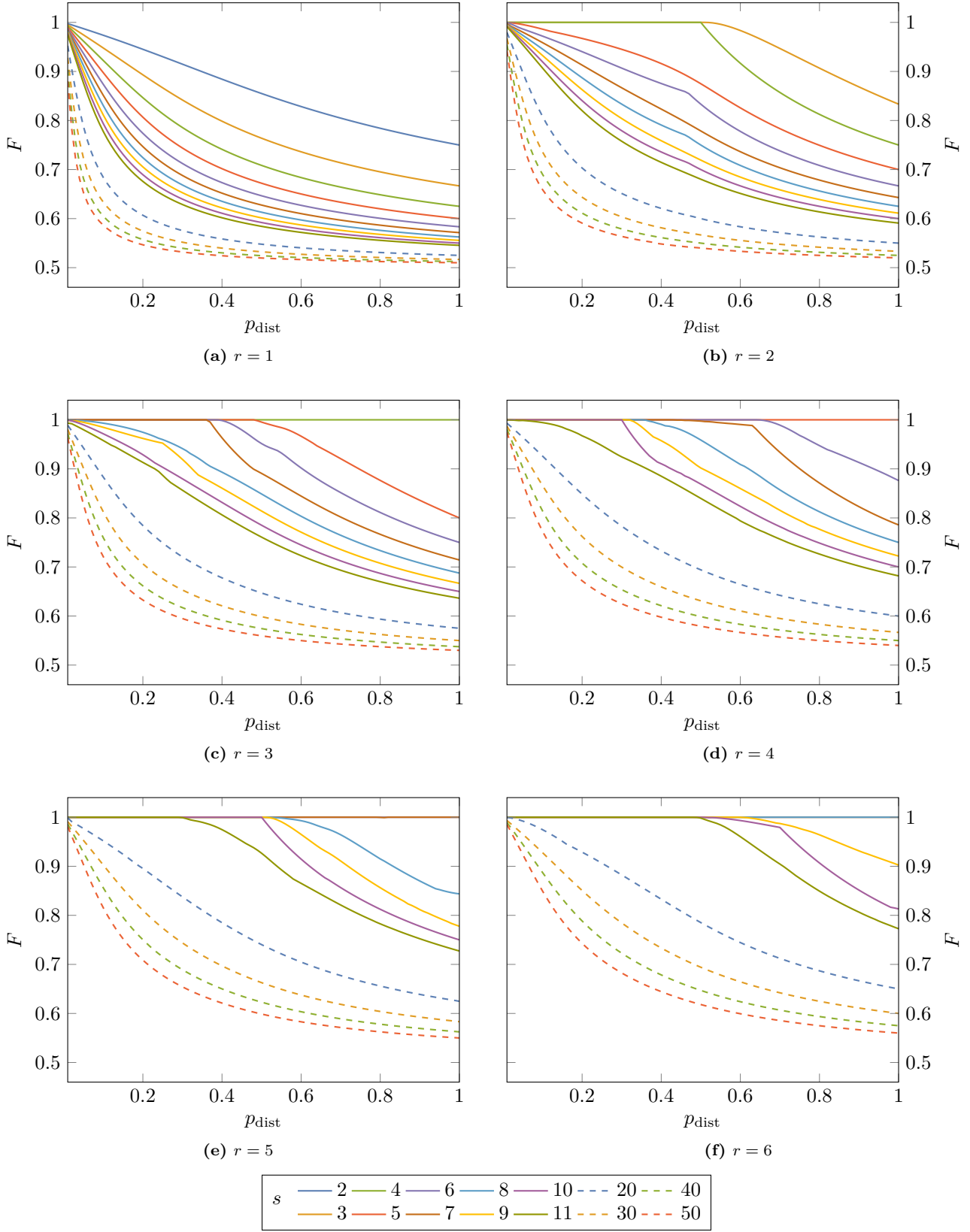
            \end{minipage}
            \clearpage
            \twocolumngrid
            \addtocounter{figure}{2}%
         }
         \begin{figure*}[t!]
            \centering%
            \begin{tikzpicture}
               \begin{axis}[xlabel=$p_{\mathrm{dist}}$, ylabel=$F$, legend pos=south west, enlarge x limits=false, axis on top, ymin=.68, ymax=1.04]
                  \addlegendimage{empty legend}
                  \addlegendentry{$(s, r)$}
                  \addplot[no marks, mpc0] table {./results/2-2-1.dat};
                  \addplot[no marks, mpc0, dashed, forget plot] table {./results/3-2-1.dat};
                  \addlegendentry{$(2, 1)$}
                  \addplot[no marks, mpc1] table {./results/2-3-2.dat};
                  \addplot[no marks, mpc1, dashed, forget plot] table {./results/3-3-2.dat};
                  \addlegendentry{$(3, 2)$}
                  \addplot[no marks, mpc2] table {./results/2-5-2.dat};
                  \addplot[no marks, mpc2, dashed, forget plot] table {./results/3-5-2.dat};
                  \addplot[no marks, mpc2, densely dash dot, forget plot] table {./results/4-5-2.dat};
                  \addlegendentry{$(5, 2)$}
               \end{axis}
            \end{tikzpicture}\hfill
            \begin{tikzpicture}
               \begin{axis}[xlabel=$p_{\mathrm{dist}}$, ylabel=$F$, legend pos=south west, enlarge x limits=false, axis on top, xmin=.35, ymin=.76, ymax=1.04, yticklabel pos=right]
                  \addlegendimage{empty legend}
                  \addlegendentry{$(s, r)$}
                  \addplot[no marks, mpc0] table {./results/2-5-3.dat};
                  \addplot[no marks, mpc0, dashed, forget plot] table {./results/3-5-3.dat};
                  \addplot[no marks, mpc0, densely dash dot, forget plot] table {./results/4-5-3.dat};
                  \addplot[no marks, mpc0, densely dotted, forget plot] table {./results/5-5-3.dat};
                  \addlegendentry{$(5, 3)$}
                  \addplot[no marks, mpc1] table {./results/2-6-4.dat};
                  \addplot[no marks, mpc1, dashed, forget plot] table {./results/3-6-4.dat};
                  \addlegendentry{$(6, 4)$}
                  \addplot[no marks, mpc2] table {./results/2-7-4.dat};
                  \addplot[no marks, mpc2, dashed, forget plot] table {./results/3-7-4.dat};
                  \addlegendentry{$(7, 4)$}
               \end{axis}
            \end{tikzpicture}
            \stepcounter{figure}%
            \caption{%
               Comparison between lower bounds on the fidelities for qubit channels (\emph{solid lines}, as in \cref{fig:fidsr}), qutrit channels (\emph{dashed lines}), ququart channels (\emph{dashed--dotted lines}), and ququint channels (\emph{dotted lines}).
               Various combinations of $s$ and $r$ are color\hyp coded.
               We only show a channel of higher dimensionality in those cases in which we have been able to find an improved lower bound numerically.
               Increasing the tested dimension may tighten the bound further.
            }
            \label{fig:fiddit}
         \end{figure*}
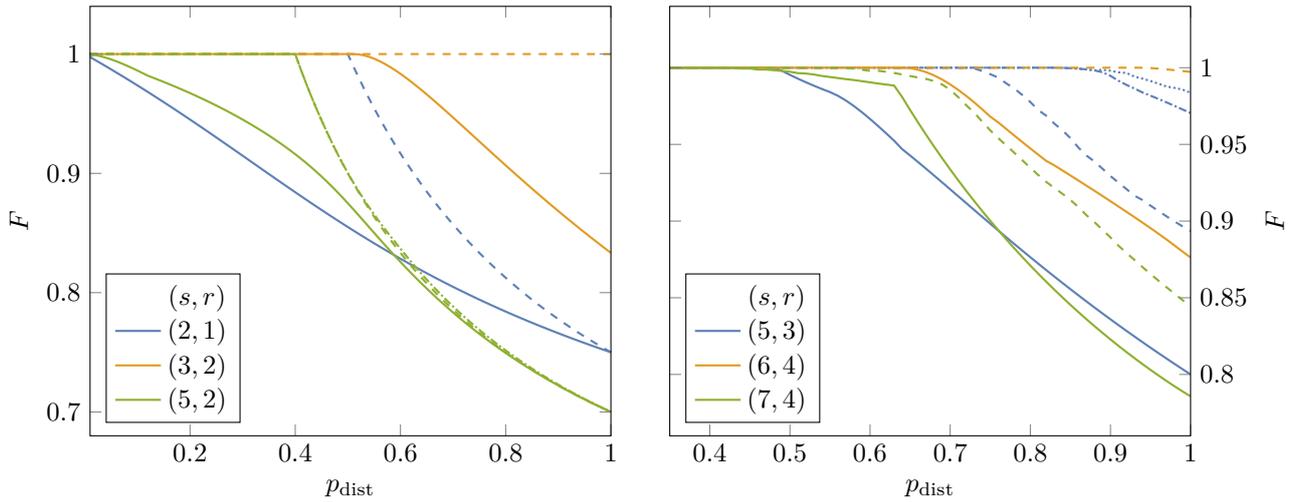%

         Our procedure allows us to simultaneously obtain optimal states and maps in order to circumvent channel attenuation.
         It is interesting to see how our results compare to known entanglement distillation schemes.
         Since these schemes typically favor a different approach, namely starting from a completely unknown state, the most fair way of conducting such a comparison is to take a known entanglement distillation protocol from the literature and apply \cref{eqn:noncon1} on it---i.e., we allow to make the most out of this protocol by choosing the optimal input state.
         We go a step further than what we did with our own schemes and allow arbitrary---in particular, also complex\hyp valued---initial states for the known protocols.
         This can now easily be done, as the optimizations are semidefinite programs.

         Still, it has to be taken into account that the protocols we are referring to actually are recurrence protocols, so they are meant to be applied multiple times in a row---typically with dramatic consequences for the success probability and state consumption.
         While by a sequential subselection instead of a parallel one or incorporating further feedback~\cite{bib:Xu2012}, their input state consumption may be greatly reduced, this is impossible without backwards communication or even multiple rounds of two\hyp way communication.
         Here, we only take into account a single application of the distillation procedure.

         \subsubsection{BBPSSW/DEJMPS}
            The first known entanglement distillation protocol was the one by Bennett \emph{et al.}~\cite{bib:Bennett1996b,*bib:Bennett1996b2}, known as BBPSSW after its authors.
            It is the predecessor of the Quantum Privacy Amplification protocol DEJMPS, by Deutsch \emph{et al.}~\cite{bib:Deutsch1996,*bib:Deutsch1998,bib:Macchiavello1998}; however, if the distillation operation is applied only once, both are equivalent.
            We depict the procedure in circuit form in \cref{fig:bbpssw1}; for a description of the protocol, see the figure caption.
            \Cref{fig:bbpssw2} shows how we adapt the protocol to our scenario for comparison.
            We compare the optimal fidelities in \cref{fig:bbpssw3} with our own schemes and can indeed observe quite significant improvements.
            Note that the DEJMPS protocol was found to be the optimal two\hyp particle distillation protocol if the shared states are Bell\hyp diagonal (as is a common assumption in entanglement distillation) and of rank at most three~\cite{bib:Rozpedek2018}, or if the possible distillation operations consist of local permutations of the Bell basis~\cite{bib:Dehaene2003}.

         \subsubsection{\texorpdfstring{\textsc{CNot}}{CNot} double selection}
            As an example of an entanglement distillation protocol among three particles, we choose the double selection circuit in~\cite{bib:Fujii2009}, which is depicted in \cref{fig:cnotcnot1}.
            Since the postselection now is on coincidences of two pairs of measurements, it is no longer possible to prepare an initial state in such a way that a successful outcome is always obtained, but instead $p_{\mathrm{dist}} \leq \frac23$.
            As is evidenced in \cref{fig:cnotcnot3}, also the double selection circuit is by far inferior to our results; in fact, we are even able to numerically arrive at a code that deterministically protects against the loss.

         \subsubsection{Further schemes}
            We did similar comparisons with the \textsc{CCNot} distillation scheme~\cite{bib:Metwally2006} ($r = 3$), the exemplary scheme for $r = 4$ by Dehaene \emph{et al.}~\cite{bib:Dehaene2003}, as well as the triple selection scheme ($r = 4$) given in~\cite{bib:Krastanov2019}.
            Since all of those procedures are based on coincidence measurements of $r - 1$ particles, the maximum success probability has an upper bound that decreases with $r$.
            The fidelity achievable by our codes naturally was always an upper bound to those schemes; and unless they are able to deliver unit fidelity results, there actually is a large gap.

   \begin{figure*}[t!]
      \centering
      \makeatletter
      \begin{tikzpicture}
         \begin{axis}[xlabel=$p_{\mathrm{tot}}$\vphantom{$L$(}, ylabel=$F$, axis on top, enlarge x limits=false, xmin=0]
            \foreach \l/\opt [count=\i from 0] in {
               1/{pos=0, anchor=310},
               11/{pos=0, anchor=260},
               21/{pos=.07, anchor=south, sloped},
               31/{pos=.05, anchor=south, sloped},
               41/{pos=.05, anchor=south, sloped},
               51/{pos=.3, anchor=south, rotate=-40},
               61/{pos=.33, anchor=south, rotate=-40},
               71/{pos=.34, above=-1pt, rotate=-40},
               81/{pos=1, anchor=north},
               91/{pos=.6, anchor=north}
            } {
               \edef\cmd{
                  \noexpand\addplot[mpc\i, no marks] table {results/bestdist\l.dat} node[\expandafter\@firstofone\opt] {$\noexpand\u{\l}{km}$};
               }
               \cmd
            }
         \end{axis}
      \end{tikzpicture}\hfill%
      \begin{tikzpicture}
         \begin{axis}[xlabel=$L$ ($\mathrm{km}$)\vphantom{$p$}, ylabel=$F$, axis on top, enlarge x limits=false, yticklabel pos=right]
            \foreach \p/\opt [count=\i from 0] in {
               5/{pos=.05, anchor=south, sloped},
               15/{pos=.03, anchor=south, rotate=-70},
               25/{pos=.03, anchor=south, rotate=-62},
               35/{pos=.03, anchor=south, sloped},
               45/{pos=.03, anchor=south, sloped},
               55/{pos=0, anchor=160},
               65/{pos=0, anchor=260},
               75/{pos=1, anchor=120},
               85/{pos=1, anchor=70},
               95/{pos=.25, anchor=north, sloped}
            } {
               \edef\cmd{
                  \noexpand\addplot[mpc\i, no marks] table {results/bestprob\p.dat} node[\expandafter\@firstofone\opt] {$\noexpand\u{\p}{\%}$};
               }
               \cmd
            }
         \end{axis}
      \end{tikzpicture}
      \addtocounter{figure}{1}%
      \caption{%
         Lower bounds on the maximally achievable fidelities for a pure loss channel.
         $p_{\mathrm{tot}}$ represents the product of the transmission with the distillation success probability.
         The transmission probabilities are based on \cref{eqn:attenuation} and the typical attenuation coefficient $\alpha = \uf{0.2}{dB}{km}$, for various distances $L$ (\emph{left}) and total success probabilities (\emph{right}), as indicated by the line labels.
         The data was obtained by taking the best possible distillation protocol according to \cref{eqn:blpsmall} for a certain probability (floored to integer percent values) out of all simulations.
         The lines are cut off at the left at the point at which direct transmission without any multiplexing (giving $F = 1$) would outperform all encodings that we studied.
         The lines are cut off at the right due to the fact that we only scanned a finite set of configurations, so that we did not consider every possible success probability---though of course, every desired probability can be achieved.
      }
      \label{fig:distdistance}
      \addtocounter{figure}{-2}%
   \end{figure*}

         \afterpage{%
            \edef\colwd{\the\columnwidth}%
            \addtocounter{figure}{-1}%
            \onecolumngrid
            \noindent\begin{minipage}[c][\textheight][c]{\linewidth}
               \begin{figure}[H]
                  \makeatletter
                  \begin{minipage}{\colwd}
                     \subfloat[\label{fig:bbpssw1}Original entanglement distillation procedure~\cite{bib:Bennett1996b,*bib:Bennett1996b2} (where we dropped the final twirling, as it is irrelevant in our case).
                        Both Alice and Bob share two entangled states (indicated by the wavy line), which, through a local twirling, they cast into Werner form (for DEJMPS, this step is replaced single\hyp qubit unitaries).
                        They apply local \textsc{CNot} operations, measure the target qubits and postselect on coincidence.]{%
                        \parbox{\linewidth}{\begin{tikzpicture}
                           \begin{yquant}
                              qubit {} alice[2];
                              qubit {} bob[2];

                              init {Alice} (alice);
                              init {Bob} (bob);

                              correlate (alice[0], bob[0]);
                              correlate (alice[1], bob[1]);

                              box {twirling} -;

                              cnot alice[1] | alice[0];
                              measure alice[1];
                              cnot bob[1] | bob[0];
                              measure bob[1];

                              box {coincidence\\postselection} (-);
                              discard alice[1], bob[1];
                           \end{yquant}
                        \end{tikzpicture}}%
                     }

                     \subfloat[\label{fig:bbpssw2}Adapted entanglement distillation scheme used for comparison.
                        Since we optimize over the initial state, we can drop all initial single\hyp qubit operations; they are automatically accounted for by the optimization.
                        The channel then suffers loss (with equal probability for all possible loss configurations) and we apply the remaining distillation operations.
                        The final postselection is now on one of the two possible outcomes, as we can imagine the measurement at Alice's site, together with her distillation procedure, to already have taken place in the preparation step.]{%
                        \parbox{\linewidth}{\hspace{.5\linewidth}\begin{tikzpicture}
                           \begin{yquant}
                              qubit {Alice} alice;
                              qubit {} send[4];
                              init {$s$} (send);

                              box {optimal\\preparation} (-);
                              box {loss} (send);
                              discard send[2-];

                              cnot send[1] | send[0];
                              measure send[1];

                              box {postcondition} (send[-1]);
                              discard send[1];
                              output {Bob} send[0];
                           \end{yquant}
                           \global\pgf@picminx=.5\dimexpr\pgf@picminx+\pgf@picmaxx\relax %
                           \global\pgf@picmaxx=\pgf@picminx
                        \end{tikzpicture}\hspace{.5\linewidth}}%
                     }%
                  \end{minipage}\hfill%
                  \begin{minipage}{\colwd}%
                     \subfloat[\label{fig:bbpssw3}Fidelities achievable through channel loss for various values of $s$ when optimized over the initial state. The solid line refers to the adapted BBPSSW scheme; the dashed line corresponds to our data as in \cref{fig:fidsr}.]{%
                        \parbox{\linewidth}{\begin{tikzpicture}
                           \begin{axis}[xlabel=$p_{\mathrm{dist}}$, ylabel=$F$, legend style={fill=none}, legend pos=south west, enlarge x limits=false, axis on top, ymin=.46, ymax=1.04, legend columns=2, legend cell align=right]
                              \addlegendimage{empty legend}
                              \addlegendentry{$s$\vphantom T}
                              \addlegendimage{empty legend}
                              \addlegendentry{}
                              \foreach \s in {3, ..., 10} {
                                 \edef\cmd{%
                                    \noexpand\addplot[no marks, mpc\the\numexpr\s-3\relax, thick] table {./Presentation-BBPSSW\s.dat};
                                    \noexpand\addlegendentry{$\s$}
                                    \noexpand\addplot[no marks, mpc\the\numexpr\s-3\relax, thick, dashed, forget plot] table {./results/2-\s-2.dat};
                                 }
                                 \cmd%
                              }
                           \end{axis}
                        \end{tikzpicture}}%
                     }%
                  \end{minipage}
                  \caption{BBPSSW/DEJMPS distillation scheme.}
                  \label{fig:bbpssw}
               \end{figure}

               \vfill
               \begin{figure}[H]
                  \makeatletter
                  \begin{minipage}{\colwd}
                     \subfloat[\label{fig:cnotcnot1}Original entanglement distillation procedure~\cite{bib:Fujii2009}.]{%
                        \parbox{\linewidth}{\begin{tikzpicture}
                           \begin{yquant}
                              qubit {} alice[3];
                              qubit {} bob[3];

                              init {Alice} (alice);
                              init {Bob} (bob);

                              correlate (alice[0], bob[0]);
                              correlate (alice[1], bob[1]);
                              correlate (alice[2], bob[2]);

                              align -;
                              cnot alice[1] | alice[0];
                              cnot alice[1] | alice[2];
                              cnot bob[1] | bob[0];
                              cnot bob[1] | bob[2];
                              h alice[2];
                              h bob[2];

                              measure alice[1-], bob[1-];
                              box {coincidence\\postselection} (-);
                              discard alice[1-], bob[1-];
                           \end{yquant}
                        \end{tikzpicture}}%
                     }

                     \subfloat[\label{fig:cnotcnot2}Adapted entanglement distillation scheme used for comparison.]{%
                        \parbox{\linewidth}{\hspace{.5\linewidth}\begin{tikzpicture}
                           \begin{yquant}
                              qubit {Alice} alice;
                              qubit {} send[5];
                              init {$s$} (send);

                              box {optimal\\preparation} (-);
                              box {loss} (send);
                              discard send[3-];

                              cnot send[1] | send[0];
                              cnot send[1] | send[2];
                              h send[2];
                              measure send[1-2];
                              box {postcondition} (send[-2]);
                              discard send[1-2];
                              output {Bob} send[0];
                           \end{yquant}
                           \global\pgf@picminx=.5\dimexpr\pgf@picminx+\pgf@picmaxx\relax %
                           \global\pgf@picmaxx=\pgf@picminx
                        \end{tikzpicture}\hspace{.5\linewidth}}%
                     }%
                  \end{minipage}\hfill%
                  \begin{minipage}{\colwd}
                     \subfloat[\label{fig:cnotcnot3}Fidelities achievable through channel loss for various values of $s$ when optimized over the initial state. The solid line refers to the adapted \textsc{CNot} double selection scheme; the dashed line corresponds to our data as in \cref{fig:fidsr}.]{%
                        \parbox{\linewidth}{\begin{tikzpicture}
                           \begin{axis}[xlabel=$p_{\mathrm{dist}}$, ylabel=$F$, legend style={fill=none}, legend pos=south west, enlarge x limits=false, axis on top, ymin=.46, ymax=1.04, legend cell align=right]
                              \addlegendimage{empty legend}
                              \addlegendentry{$s$\vphantom T}
                              \foreach \s in {4, ..., 10} {
                                 \edef\cmd{%
                                    \noexpand\addplot[no marks, mpc\the\numexpr\s-3\relax, thick] table {./Presentation-CNotCNot\s.dat};
                                    \noexpand\addlegendentry{$\s$}
                                    \noexpand\addplot[no marks, mpc\the\numexpr\s-3\relax, thick, dashed, forget plot] table {./results/2-\s-3.dat};
                                 }
                                 \cmd%
                              }
                           \end{axis}
                        \end{tikzpicture}}%
                     }%
                  \end{minipage}
                  \caption{\textsc{CNot} double selection distillation scheme.}
                  \label{fig:cnotcnot}
               \end{figure}
            \end{minipage}
            \clearpage
            \twocolumngrid
            \addtocounter{figure}{1}%
         }%

   \section{Discussion}\label{sec:disc}
      \subsection{Taking attenuation into account}
         All results on entanglement distillation protocols---except from \cref{eqn:blpmulrecv}---cannot be immediately related to the task of counteracting channel losses.
         The abscissa in \cref{fig:fidsr} refers to the success probability of the distillation scheme \emph{alone}.
         We still have to take into consideration the probability of successfully transmitting the required multiplexed state, $P_{\mathrm{trans}} = \sum_{i = r}^s \binom si p_{\mathrm{trans}}^i \bar p_{\mathrm{trans}}^{s - i}$.
         In order to check this, we assembled the numerical data from all simulations that we carried out and replaced the success probability of distillation by the total success probability.
         We then took the best fidelity possible for any given probability and plotted this data in \cref{fig:distdistance}.

         We can indeed observe how our optimization schemes are able to deliver entangled states in the regimes at which direct transmission fails.
         The fidelity of these states of course depends strongly on the desired total success probability and is thus tunable.

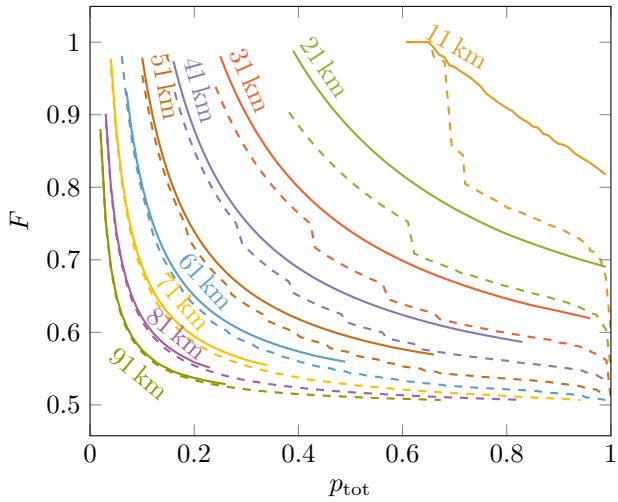
\begin{figure}[t!]
            \centering
            \begin{tikzpicture}
               \begin{axis}[xlabel=$p_{\mathrm{tot}}$, ylabel=$F$, legend pos=outer north east, axis on top, legend cell align=right, enlarge x limits=false, xmin=0]
%                  \addlegendimage{empty legend}
%                  \addlegendentry{$\u{L}{\hphantom{km}}$}
                  \foreach \l/\opt [count=\i from 1] in {
                     11/{pos=.2, anchor=south, rotate=-30},
                     21/{pos=.07, anchor=south, sloped},
                     31/{pos=.06, anchor=south, sloped},
                     41/{pos=.06, anchor=south, sloped},
                     51/{pos=.06, anchor=south, sloped},
                     61/{pos=.5, anchor=south, sloped},
                     71/{pos=.65, anchor=south, sloped},
                     81/{pos=.8, above=-1pt, sloped},
                     91/{pos=.7, anchor=north, sloped}
                  } {
                     \edef\cmd{
                        \noexpand\addplot[mpc\i, no marks] table {fullR/bestdist\l.dat} node[\expandafter\@firstofone\opt] {$\noexpand\u{\l}{km}$};
                        \noexpand\addplot[mpc\i, no marks, dashed, forget plot] table {results/bestdist\l.dat};
%                        \noexpand\addlegendentry{$\noexpand\u{\l}{km}$}
                     }
                     \cmd
                  }
               \end{axis}
            \end{tikzpicture}
            \caption{%
               Lower bounds on the maximally achievable fidelities for a pure loss channel, using maps that depend on the number of arrived states.
               The transmission probabilities are based on \cref{eqn:attenuation} and the typical attenuation coefficient $\alpha = \uf{0.2}{dB}{km}$ for various distances~$L$.
               The \emph{dashed lines} are the same as in \cref{fig:distdistance}, where we only considered a fixed map; the \emph{solid lines} allow for different maps depending on the actual number of received states.
               The data was obtained by taking the best possible distillation protocol according to \cref{eqn:blpmulrecv} for a certain probability (floored to integer percent values) out of all simulations.
               The lines are cut off at the left at the point at which direct transmission without any multiplexing (giving $F = 1$) would outperform all encodings that we studied.
               The lines are cut off at the right due to the fact that we only scanned a finite set of configurations, so that we did not consider every possible success probability---though of course, every desired probability can be achieved.
               Note that there were much fewer samples from which we took the optimum for the solid lines compared to the dashed ones.
            }
            \label{fig:distdistancesmul}
         \end{figure}

         We note that there is potential for improvement by using \cref{eqn:blpmulrecv}, i.e., employing different maps depending on the different possible values of~$r$.
         We considered this \namecref{eqn:blpmulrecv} for a subset of the previous configurations.
         \Cref{fig:distdistancesmul} shows that this can indeed lead to a large payoff, in particular when high success probabilities are desired: despite our pool of protocols being much smaller, the fidelities obtained in this way in some instances are almost as good as the previous ones for $\u{10}{km}$ less distance (almost $\u{60}{\%}$ higher success probability).

      \subsection{Implications for conventional entanglement distillation}\label{sec:discentdist}
         We will use this \namecref{sec:discentdist} to slightly deviate from our original goal to find suitable procedures that work in a quantum repeater scenario and consider the task of ``pure'' entanglement distillation instead.
         Recurrence schemes have the advantage of working in the low\hyp fidelity regime, but they have exponentially low yields and require two\hyp way communication.
         Therefore, usually, a recurrence distillation is carried out only until the fidelity is high enough to allow the hashing protocol~\cite{bib:Bennett1996} to continue, which can asymptotically distill perfect states using one\hyp way communication only.

         Though not designed with this particular purpose in mind, our protocols are of course valid ``pre\hyp hashing'' schemes.
         To investigate this, we calculate the von~Neumann entropy~$S$~\cite{bib:vonNeumann2018} of the final two\hyp party state~$\rho_{\mathrm f}$ after a complete dephasing in the Bell basis.
         After this single step of distillation, we can, using only one\hyp way communication, asymptotically distill perfect Bell pairs with a rate $1 - S$~\cite{bib:Bennett1996}.
         In fact, for this particular use, it would have been a meaningful alternative not to maximize~$F$, but instead minimize~$S$ in all previous programs.
         However, while the entropy of a state after a dephasing in a particular basis can be accessed in a convex program by means of the exponential cone~\cite{bib:Mosek2021}, being concave, it cannot be minimized efficiently.
         We therefore use our fidelity approach as an approximation to the entropy optimization, where we calculate the latter \emph{a posteriori}.

\begin{figure}[t!]
            \centering
            \begin{tikzpicture}
               \begin{axis}[xlabel=$L$ ($\mathrm{km}$), ylabel=$S$, legend pos=south east, axis on top, legend cell align=left, enlarge x limits=false, extra y ticks={1}, extra y tick style={grid=major, tick label style={draw=none}}, xtick={0, 20, ..., 80}]
                  \addplot[mpc0, no marks] table {results/entrdistdirect.dat};
                  \addlegendentry{direct transmission}
                  \addplot[mpc1, no marks] table {results/entrdistmult.dat};
                  \addlegendentry{optimized distillation}
                  \addplot[mpc2, no marks] table {results/entrdist5qubit.dat};
                  \addlegendentry{$\code{5, 1, 3}$ code (full)}
                  \addplot[mpc3, no marks] table {results/entrdist5qubitred.dat};
                  \addlegendentry{$\code{5, 1, 3}$ code (red.)}
                  \node[pin=30:{$(3, 6, 4, 1)$}] at (3, 0.217839) {};
                  \node[pin=10:{$(4, 5, 3, 1)$}] at (7, 0.739908) {};
                  \node[pin={[pin distance=4ex]25:{$(2, 48, 1, 1)$}}] at (11, 1.14555) {};
                  \node[pin=175:{$(2, 72, 1, 1)$}] at (90, 1.66229) {};
               \end{axis}
            \end{tikzpicture}
            \caption{%
               Entropies $S$ of the Bell\hyp diagonalized normalized final states $\hat\rho_{\mathrm f}$ after transmission through a pure loss channel.
               The transmission probabilities are based on \cref{eqn:attenuation} and the typical attenuation coefficient $\alpha = \uf{0.2}{dB}{km}$, for various distances $L$.
               More precisely, as we rely on one\hyp way communication only, we look at the von~Neumann entropies of $p_{\mathrm{tot}} \hat\rho_{\mathrm f} + (1 - p_{\mathrm{tot}}) \openone/4$.
               The data was obtained by taking the best possible distillation protocol according to \cref{eqn:blpsmall} out of all simulations at each sample point (integer kilometers); the multiplexed curve is therefore an upper bound.
               For selected points, we give the combinations $(d, s, r, p_{\mathrm{dist}})$ that gave rise to the optimal value.
               The largest sampled multiplexing value was $s = 75$.
               We also compare this with the entropies obtained by using the perfect $\code{5, 1, 3}$ code~\cite{bib:Laflamme1996}, where we take optimal correction maps that can either depend on the full information of which particular configuration arrived, or only take into account the reduced information of how many qubits survived.
            }
            \label{fig:compent}
         \end{figure}
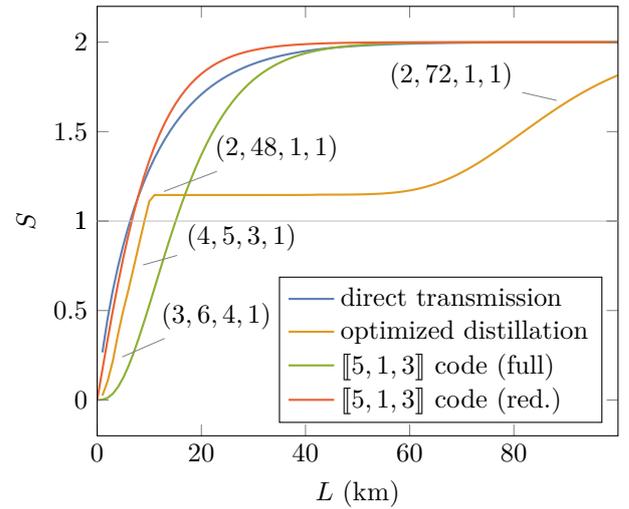

         There are two main ways in which we might connect our results with a hashing protocol.
         The one\hyp way communication approach requires us to mix the final state $\rho_{\mathrm f}$ with a maximally mixed state, with weights corresponding to the total success and failure probability, respectively.
         While this procedure always gives a state, the fidelities that can be achieved are naturally rather modest; this is only viable for relatively short distances and low\hyp loss scenarios.
         Naturally, the optimal distillation protocol will always be the deterministic one, as the maximally mixed state that compensates for the reduced success probability is worse than the benefit that we can possibly obtain from an increased fidelity.
         When we compare the entropy of directly transmitted states with the entropy after the best of our distillation schemes in \cref{fig:compent}, we are still able to find significant improvements.
         The distillation can increase the critical distance with one\hyp way communication from $\u{6}{km}$ to about $\u{9}{km}$.
         Compared to QECCs, which can offer up to $\u{15}{km}$---and indeed, the perfect $\code{5, 1, 3}$ code~\cite{bib:Laflamme1996} accomplishes this---this seems to suggest far from optimal results.
         However, in \cref{fig:compent}, we depict the performance of the $\code{5, 1, 3}$ code\footnote{%
            To make the comparison most meaningful, we take the logical states of the code to form the encoded state $\ket0 \otimes \ket0_{\mathrm L} + \ket1 \otimes \ket1_{\mathrm L}$.
            For \emph{every} loss configuration---which is much more than what we allow for in our codes---we optimize to find the ideal map that gives a unit\hyp trace final state with maximal fidelity.%
         }, and we also investigate how well this code can perform if applied to our chosen setting, where we do not fully exploit the information on which particular loss configuration arrived.
         In the latter case, we clearly observe how our codes surpass the QECC---despite that fact that for our codes, we even fix~$r$, while we still allow $r$\hyp dependent correction maps for the QECC.
         We conjecture that indeed, the possibility to discriminate the basis states with the same number of set bits is crucial to get the best possible performance.

         For our codes, we observe that only a few combinations $(d, s, r, p_{\mathrm{dist}})$ turn out to be optimal protocols; with less than $\u{50}{\%}$ losses, these are particularly promising qudit protocols; for high losses, we naturally have $r = 1$.
         It is interesting to see that the best protocol for a long time is with $s = 48$, before it switches to $s = 73$, where $s = 75$ was the maximum simulated.
         Only looking at the fidelity---which is monotonically decreasing with $s$---there is no particular reason to single out these two combinations; however, this is where the difference between fidelity and entropy optimization shows.
         A multitude of different entropies can give the same fidelity; and due to the randomness of our starting points in the optimizations, the $s \in \{ 48, 73 \}$ configurations turned out to give the best entropy.
         This already shows the limitations of applying our single\hyp shot scenario in the asymptotic regime; it is highly likely that the upper bound in \cref{fig:compent} can in fact be lowered by more suitable optimization strategies or exploiting even higher\hyp dimensional carriers.
         Still, this scenario is not suitable for surpassing the $\u{15}{km}$ barrier; it aims at distilling perfect states and relies on deterministic protocols, so that it is bound by the no\hyp cloning theorem.

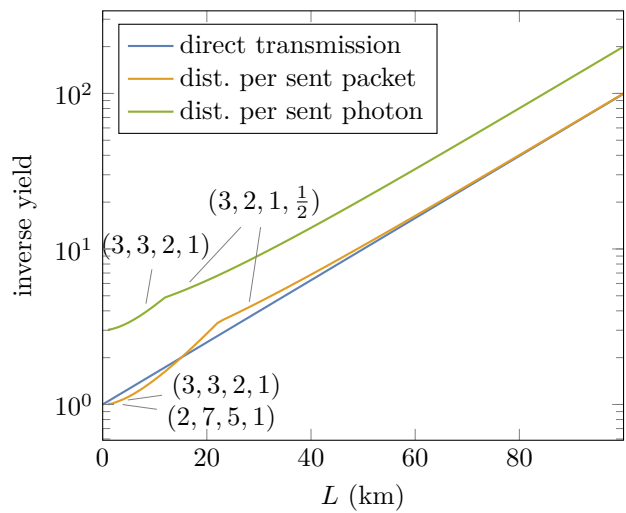
\begin{figure}[b!]
            \centering
            \begin{tikzpicture}
               \pgfplotsset{set layers=axis on top}
               \begin{axis}[xlabel=$L$ ($\mathrm{km}$), ylabel=inverse yield, legend pos=north west, axis on top, legend cell align=left, enlarge x limits=false, ymode=log, xtick={0, 20, ..., 80}]
                  \addplot[mpc0, no marks, domain=0:100] {exp(.046*x)};
                  \addlegendentry{direct transmission}
                  \addplot[mpc1, no marks] table {results/invyieldpacket.dat};
                  \addlegendentry{dist. per sent packet}
                  \addplot[mpc2, no marks] table {results/invyield.dat};
                  \addlegendentry{dist. per sent photon}
                  \node[pin={[xshift=1ex]90:{$(3, 3, 2, 1)$}}] at (8, 3.876) {};
                  \node[pin={[pin distance=5ex, name=pin]75:{$(3, 2, 1, \frac12)$}}] at (15, 5.32231) {};
                  \node[pin={[yshift=1ex]-5:{$(2, 7, 5, 1)$}}] at (2, 1.01846) {};
                  \node[pin={[yshift=-1ex]5:{$(3, 3, 2, 1)$}}] at (3, 1.04774) {};
                  \draw[every pin edge, shorten <= .5ex] (28, 4.20563) -- (pin);
               \end{axis}
            \end{tikzpicture}
            \caption{%
               Inverse yields for the distillation of perfect Bell pairs across a loss channel of given length.
               The data was obtained by taking the best possible distillation protocol according to \cref{eqn:blpsmall} out of all simulations at each sample point (integer kilometers); the multiplexed curve is therefore an upper bound.
               For selected points, we give the combinations $(d, s, r, p_{\mathrm{dist}})$ that gave rise to the optimal value.
               \emph{Direct transmission} corresponds to the inverse of the success probability of transmission.
               The curve \emph{dist. per sent packet} assumes that each instance of our protocol counts as one channel use; the curve \emph{dist. per sent photon} counts every protocol use by its number of sent photons~$s$.
               Note that since this weighting affects which protocol is optimal, both curves do in general not give the same protocol for the same~$L$ (although they do over a large range).
            }
            \label{fig:yield}
         \end{figure}

         The second way to combine our results with hashing is to consider a distillation task that allows for two\hyp way communication.
         In this way, both the unsuccessful transmission as well as an unsuccessful distillation can be signaled back and therefore do not have an impact on the state, at the expense of being more resource\hyp intensive.
         The interesting quantity in this regard is the (inverse) yield: How many packets of $s$~particles are required, on average, to obtain a single perfect Bell pair, when we choose our best distillation protocol and feed its output into the hashing protocol?
         In \cref{fig:yield}, we plot the inverse yields, both when we count the application of our protocol as a single send step and when we count it in terms of the number of sent particles~$s$.
         We can observe that for short distances, $p_{\mathrm{dist}} = 1$ is still optimal; but for longer distances, $p_{\mathrm{dist}} = \frac12$ in fact becomes the best protocol.
         In all cases, the best protocols are the ones that already give unit fidelity, i.e., those that completely avoid the hashing protocol.
         \Cref{fig:yield} shows that, when measured in terms of packets, we get a clear advantage in the $p_{\mathrm{trans}} > \frac12$ range, but direct transmission---where we send Bell state halves---always has the advantage in the high\hyp loss regime.

         We remark that in the two\hyp way communication scenario, it is meaningful to ask about multiple successive applications of our protocols.
         Without the possibility to signal back, this would be no better---but potentially worse---than one of our protocols with twice the number of received particles, while feedback can in principle allow to increase the yield.
         This comes at the expense of delaying the distillation operation at the sender's site until the feedback has arrived.

      \subsection{Implications for quantum key distribution}
         Our procedure was motivated by the problem of quantum key distribution, where the final secret key rate is linearly dependent on $p_{\mathrm{trans}}$, but can drop to zero if other errors are too high.
         Note that here, we allow for two\hyp way communication, as long as it has to occur only between the communication partners at the end of the line and not between repeaters; hence, it only adds an initial latency, but does not affect scalability.

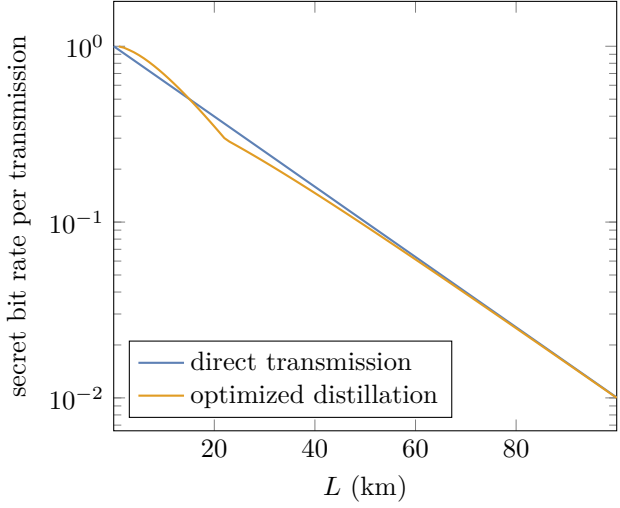
\begin{figure}[htbp]
            \centering
            \begin{tikzpicture}
               \begin{axis}[xlabel=$L$ ($\mathrm{km}$), ylabel=secret bit rate per transmission, legend cell align=left, ymode=log, legend pos=south west, enlarge x limits=false, xtick={20, 40, 60, 80}, height=\axisdefaultheight, width=8.2cm, ymin=.0065, ymax=1.8]
                  \addplot[mpc0, no marks, domain=0:100] {exp(-.046*x)};
                  \addlegendentry{direct transmission}
                  \addplot[mpc1, no marks] table {results/skrate.dat};
                  \addlegendentry{optimized distillation}
               \end{axis}
            \end{tikzpicture}
            \caption{%
               Secret bit rates per sent packet after transmission through a pure loss channel, achievable via two\hyp way communication between the two stations.
               The transmission probabilities are based on \cref{eqn:attenuation} and the typical attenuation coefficient $\alpha = \uf{0.2}{dB}{km}$, for various distances $L$.
               The data was obtained by taking the best possible distillation protocol according to \cref{eqn:blpsmall} out of all simulations at each sample point (integer kilometers); the multiplexed curve is therefore a lower bound.%
            }
            \label{fig:qkd}
         \end{figure}

         For a repeaterless operation mode, we plot the secret key rate, \cref{eqn:qkdrate}, in \cref{fig:qkd} and again see significant rate improvements for short lengths (where the optimal protocols are deterministic), while direct transmission is at the advantage in the high\hyp loss regime (where the optimal protocols are probabilistic---using \cref{eqn:blpmulrecv}, they even give the same rates as the direct transmission, but cannot surpass them).

         We will now apply our protocol in the scenario it was developed for, namely, for the use of a quantum repeater.
         We take the simplest case and assume that we use a single quantum repeater.
         For all of our protocols, we carry out the following procedure.
         Alice and the first repeater individually prepare a copy of the multi\hyp qubit state as dictated by the protocol; the repeater sends the appropriate parts to Bob, Alice sends them to the repeater.
         Both the first repeater and Bob carry out their distillation procedure as required by the protocol; the global state now is $\rho_{\mathrm f}^{\mathrm{A R}} \otimes \rho_{\mathrm f}^{\mathrm{R B}}$, where we denoted the systems in superscripts, and $\rho_{\mathrm f}$ is precisely the best normalized approximation of $\ketbra{\Phi^+}{\Phi^+}$ that the protocol was able to deliver.
         The repeater now carries out an optimal entanglement swapping protocol for this particular state.
         This protocol is defined by the following convex optimization problem (in fact, the objective is not convex, but is easily obtained from two convex optimizations):
         \begin{equation}
            \def\pl{\mathord+}\def\mi{\mathord-}
            \mathclap{\left\{\begin{aligned}
               \MoveEqLeft \min_{\{ \Lambda_k, \rho_k \}_{k = 1}^4, e_X, e_Z} \{ e_X, e_Z \} \\
               & \centerasif{=}{\text{subject to}} \\
               C(\Lambda_k)
               & \succeq 0 \ \forall k \\
               \smash[b]{\sum_k}\vphantom{\sum} \Lambda_k
               & = \openone \\
               \rho_k
               & = (\openone^{\mathrm A} \otimes \Lambda_k \otimes \openone^{\mathrm B})\bigl[
                      \rho_{\mathrm f}^{\mathrm{A R}} \otimes
                      \rho_{\mathrm f}^{\mathrm{R B}}
                   \bigr] \\
               e_X
               & \begin{multlined}[t]
                 = \braket{\pl\mi | \rho_1 + \rho_2 | \pl\mi} + \braket{\mi\pl | \rho_1 + \rho_2 | \mi\pl} \\ +
                   \braket{\pl\pl | \rho_3 + \rho_4 | \pl\pl} + \braket{\mi\mi | \rho_3 + \rho_4 | \mi\mi}
                 \end{multlined} \\
               e_Z
               & \begin{multlined}[t]
                 = \braket{01 | \rho_1 + \rho_3 | 01} + \braket{10 | \rho_1 + \rho_3 | 10} \\ +
                   \braket{00 | \rho_2 + \rho_4 | 00} + \braket{11 | \rho_2 + \rho_4 | 11}\text.
                 \end{multlined}
            \end{aligned}\right.} \label[program]{eqn:entswap}
         \end{equation}
         Here, the four possible instruments of the protocol (in standard entanglement swapping, those would be the projections onto the four Bell basis elements) are given by the maps $\Lambda_k$, which in total are trace\hyp preserving\footnote{%
            In fact, instead of trace preservation, we can again ask for probabilistic swapping maps.
            Computationally, the probability must be fixed beforehand, so this requires a line scan.
            We also investigated this case and found no improvements by a nondeterministic swap.%
         }.
         Their application leads to four (subnormalized) possible outcomes $\rho_k$, defined at Alice's and Bob's site.
         In standard entanglement swapping, we would now require local correction operations; this amounts to the fact that the bit error rates in the two bases, $e_X$ and $e_Z$, are calculated based on anticorrelations for half of the outcomes, but on correlations for the other half.

\begin{figure}[t!]
            \centering
            \begin{tikzpicture}
               \begin{axis}[xlabel=$L$ ($\mathrm{km}$), ylabel=secret bit rate per transmission, legend cell align=left, ymode=log, enlarge x limits=false, xtick={20, 40, 60, 80}, height=\axisdefaultheight, width=8.2cm, yticklabel pos=right, ymin=.0065, ymax=1.8]
                  \addplot[mpc0, no marks, domain=1:100] {exp(-.046*x)};
                  \addlegendentry{direct transmission}
                  \addplot[mpc1, no marks] table {results/skrate1rep.dat};
                  \addlegendentry{optimized distillation}
               \end{axis}
            \end{tikzpicture}
            \caption{%
               Secret bit rates per sent packet after transmission through a pure loss channel, achievable via two\hyp way communication between the two final stations, with one quantum repeater at $L/2$.
               The transmission probabilities are based on \cref{eqn:attenuation} and the typical attenuation coefficient $\alpha = \uf{0.2}{dB}{km}$, for various distances $L$.
               The data was obtained by taking the best possible distillation protocol according to \cref{eqn:blpsmall} out of all simulations, combined with optimal entanglement swapping at each sample point (integer kilometers); the multiplexed curve is therefore a lower bound.%
            }
            \label{fig:qkd1rep}
         \end{figure}
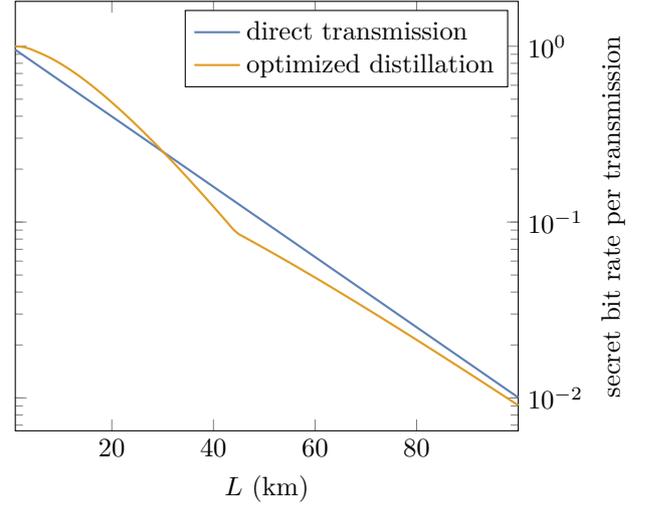

         Running \cref{eqn:entswap} for all protocols, we obtain the bit error rates with one repeater.
         We combine them with the success probabilities to get the true asymptotic secret bit rates and then compare them with the rates for direct transmission over the \emph{double} length in \cref{fig:qkd1rep}.

         While here, we still observe the same behavior that as soon as the individual stations are more than $\u{15}{km}$ apart, direct transmission is advantageous, the repeater advantage clearly shows over the full length.
         Interestingly, we find that the optimal protocol for most of the relevant range is again the deterministic $d = 3$, $s = 3$, $r = 2$, $p_{\mathrm{dist}} = 1$, which allows for perfect distillation.
         Since in this case, traditional entanglement swapping is optimal, the results can be generalized in a straightforward way to an arbitrary number of repeaters using this particular protocol.

      \subsection{Experimental remarks}
         Our scheme delivers numerically optimized states and maps.
         While analytical optimality proofs are not available, extensive numerical testing suggests that these result are close to the optimum.
         We always assumed that the channel itself is a pure loss channel that exhibits no further errors and that both state preparation and execution of the map are flawless.
         Due to these simplifications, our results are theoretical bounds.
         Real\hyp life noise will put additional constraints on the actual performance; however, having the bounds in mind allows to assess whether improvements in the quality of, say, state preparation are worthwhile.

         Both the preparation and the distillation map are very well under control and we can expect that the continuous development of quantum technology will allow to approach the idealized states and maps to a better and better degree.
         However, additional sources of error in the channel (or necessary pre- and post\hyp channels such as frequency conversion) must be thoroughly investigated.
         While our simulation methods are also able to incorporate further errors, we might instead try to nest the loss\hyp tolerant codes with other well\hyp established QECCs.
         It is not immediately obvious that this approach can work: If we first encode our data with the loss\hyp tolerant code and then each physical qubit with a traditional code, due to high losses, the traditional code will not even be able to restore the inner code.
         Hence, we must first encode our data with a traditional QECC and then with a loss\hyp tolerant code.
         Upon recovery, we must compensate for losses; but is this still possible if other errors deteriorated the state?

         We finally remark that our codes rely heavily on the use of Dicke states.
         In the last years, the preparation of qubit Dicke states has received some attention and by now, there are various theoretical proposals as well as working implementations, e.g., in ions~\cite{bib:Hume2009,bib:Ivanov2013}, atoms~\cite{bib:Stockton2004,bib:Xiao2007,bib:Shao2010}, quantum dots~\cite{bib:Zou2003}, photonic systems~\cite{bib:Prevedel2009,bib:Wieczorek2009}, or as a quantum circuit~\cite{bib:Baertschi2019}.
         Some already address how to create arbitrary superpositions of Dicke states~\cite{bib:Ivanov2013,bib:Keating2016,bib:Mu2020}.
         Most, but not all~\cite{bib:Kasture2018} are limited to relatively few qubits.
         Qudit Dicke states are much less explored; we are aware only of one very recent proposal that allows to efficiently verify qudit Dicke states~\cite{bib:Zihao2021}, which may also be a starting point when it comes to implementing the maps.

   \section{Conclusion}\label{sec:out}
      The contributions in this paper are twofold.

      First, we describe a general numerical methodology that allows to optimize quantum problems in which both a state as well as a map acting on a linearly transformed version of this state are unknowns.
      These methods are quite general in their applicability and we expect them to be able to offer helpful insight into a lot of problems that require simultaneous optimization over states and maps.
      In fact, such problem formulations are very common especially in the definition of resource theoretic measures such as entanglement~\cite{bib:Plenio2007} or channel capacities~\cite{bib:Gyongyosi2018}.
      While the limit of infinite copies that is often required in these measures is of course out of reach of a numerical algorithm, optimizing for the first couple of copies certainly help in building an intuition; and if the measure features some kind of additivity property, this allows to derive bounds.

      We then use our methodology to carry out a detailed study on how a pure loss channel can be compensated for by using new codes and procedures for entanglement distillation.
      All codes are parametrized, so that in an application, we can choose between higher fidelities or higher total success probabilities.
      While we indeed found that in some parameter ranges, non\hyp deterministic distillation procedures gave better secret bit rates, this only occurred in the high\hyp loss regime in which direct transmission would deliver results of at least the same quality.
      This does not rule out the possibility of increasing the rate by probabilistic distillation, but the restricted set of state we looked at might not be able to validate this.

      While most of our studies were carried out on qubits, we also investigated higher\hyp dimensional carriers.
      For the particular case of $p_{\mathrm{dist}} = 1$, this corresponds to qudit erasure correcting codes and can also be extended to qudit quantum error correcting codes, a subject with only few contributions so far~\cite{bib:Hostens2005,bib:Bullock2007,bib:Gheorghiu2014}.
      For the case of the reduced erasure situation, we found that sometimes, using higher\hyp dimensional information carriers instead of qubits can significantly boost the fidelity, keeping the same distillation success probability.

      We showed that further advantage can be gained if the receiver is able to choose the postprocessing map dependent on the number of arrived states.
      Our procedure was limited to the reduced erasure case, in which we exploited the erasure information only by removing failed slots from the input to the distillation map, but not by choosing different maps for different configurations.
      This approach facilitated a mapping to low\hyp dimensional vector spaces, which allowed us to carry out trustworthy simulations for a large number of qubits.
      However, we also found that the knowledge of the precise erasure configuration is actually able to significantly boost the distillation fidelity.
      Hence, the results that are optimal in our scenario can in fact be improved much further.
      While the full Hilbert spaces are hardly tractable for larger number of photons, it may be possible to use our results as a starting point and single out various smaller\hyp scaled subproblems that arise from expanding the Dicke basis, which may then in fact render the full\hyp dimensional optimization feasible.

      Our approach opens a new way to the task\hyp driven derivation of (probabilistic) error correcting or distillation codes and is able to significantly enhance the quality of signal transmission through lossy channels even under the reduced erasure regime, which does not exploit all available information.
      We expect that our methodology can be applied to design other codes---either of few qubits or exploiting symmetries---for example to protect small registers for certain operations in a quantum computer against noise and errors.
      The advantage of the optimization procedure is that a better characterization of the noise can directly manifest in better protocols---provided effective states and maps can be devised from the intended ones linearly\footnote{%
         For the convex iteration and the outer optimization, more complicated dependencies can be thought of.%
      }.

      We investigated entanglement generation as an elementary building block of QKD.
      A set of protocols for different parameters generated by numerical studies as outlined in this paper can then be used in other optimization algorithms that are concerned with scaling up single\hyp link connections to whole networks~\cite{bib:Jiang2007,bib:Goodenough2021}.

      We already pointed at direct possible sequels to this work; in particular, trying to use the full information available about the erasures should deliver much more optimized schemes.
      A second step is to take into account other typical sources of quantum errors in channels, preparation and processing, and whether a concatenation of loss\hyp tolerant with standard QECCs is able to provide sufficient protection.
      In this way, we can find optimal realizations under realistic constraints, as opposed to the theoretical limits that we considered in this paper.

   \section*{Code and data availability}
      The full code to carry out the simulations is available on GitHub~\cite{bib:Desef2021}; see \cref{fig:python} and the Readme of the repository for a brief overview.
      Numerical data is available from the authors upon request.

   \begin{acknowledgments}
      We appreciate helpful discussions with Thomas Theurer.
      This project is funded by the Federal Ministry of Education and Research~(BMBF) grants number \href{https://foerderportal.bund.de/foekat/jsp/SucheAction.do?actionMode=view&fkz=16KIS0875}{16KIS0875} and \href{https://foerderportal.bund.de/foekat/jsp/SucheAction.do?actionMode=view&fkz=16KISQ006}{16KISQ006}.
      The authors acknowledge support by the state of Baden\hyp Württemberg through bwHPC and the German Research Foundation~(DFG) through grant~no INST~40/575-1~FUGG (JUSTUS~2 cluster).
   \end{acknowledgments}

   \begingroup
   \small\RaggedRight
   \bibliography{Paper}
   \endgroup

   \appendix
   \begin{widetext}
   \section{State after partial tracing}\label{app:partracedit}
      In this appendix, we will determine the state of the system that arises from \cref{eqn:symstatedit} after tracing out $r - s$ subsystems at the receiver's site.
      To begin with, the full state is given by
      \begin{equation*}
         \rho
         = \sum_{a_1, a_2 = 0}^1 \; \sum_{\vv i, \vv j \in [s]_d}
           \rho_{a_1, a_2}(\vv i, \vv j)
           \ketbra{a_1 D^s_{\vv i}}{a_2 D^s_{\vv j}}\text,
         \tag{\ref{eqn:symstatedit}}
      \end{equation*}
      where $[s]_d$ is defined in \cref{eqn:defsd} and $\ket{D^s_{\vv i}}$ in \cref{eqn:defddicke}.
      The symmetry of the Dicke states makes it clear that it does not matter which ones of the qudits are traced out; so let us without loss of generality assume that we keep, at the receiver's site, the first $r$~qudits and trace out the remaining $s - r$~qudits.

      Let now $\vv i \in [s]_d$ be fixed.
      To simplify the notation, we define
      \begin{equation*}
         B(s, \vv i)
         \coloneqq \Bigl\{ (x_1, \dotsc, x_s) \in \{ 0, \dotsc, d -1 \}^s : \bigl\lvert \{ j : x_j = m \} \bigr\rvert = i_m \ \forall m \in \{ 0, \dotsc, d -1 \} \Bigr\}
      \end{equation*}
      as the set over all $d$-ary strings of length~$s$ in which the digit~$m$ occurs $i_m$~times.
      We can then write the definition of the Dicke state as
      \begin{equation*}
         \ket{D^s_{\vv i}} = \frac{1}{\sqrt{\binom{s}{\vv i}}}
            \sum_{\sigma \in B(s, \vv i)}
               \ket{\underbrace{\hbox to 2cm{\hfill$\sigma_1$\hfill}}_{\text{length }r}
                    \underbrace{\hbox to 2cm{\hfill$\sigma_2$\hfill}}_{\text{length }s - r}}\text,
      \end{equation*}
      splitting the digit string $\sigma$ into two parts.

      We now consider how to properly split the sum in these two parts.
      We will use the vector $\vv k \in [r]_d$ to describe how often the individual digits occur in the first $r$~digits, and $\vv k' \in [s - r]_d$ for the same in the rest.
      We have $\vv k + \vv k' = \vv i$.
      With this, we get
      \begin{align*}
         \sum_{\substack{\sigma \in B(s, \vv i) \\ \tau \in B(s, \vv j)}} \ketbra\sigma\tau
         & = \sum_{\vv k, \vv\ell \in [r]_d}
             \sum_{\vv k', \vv\ell' \in [s - r]_d}
             \delta_{\vv k + \vv k', \vv i}
             \delta_{\vv\ell + \vv\ell', \vv j}
             \sum_{\substack{\sigma_1 \in B(r, \vv k) \\ \tau_1 \in B(r, \vv\ell)}}
             \sum_{\substack{\sigma_2 \in B(s - r, \vv k') \\ \tau_2 \in B(s - r, \vv\ell')}}
                \ketbra{\sigma_1\sigma_2}{\tau_1\tau_2}\text,
         \intertext{where $\delta$ is Kronecker's delta. The partial trace can now easily be written down:}
         \tr_2 \sum_{\substack{\sigma \in B(s, \vv i) \\ \tau \in B(s, \vv j)}} \ketbra\sigma\tau
         & = \sum_{\vv k, \vv\ell \in [r]_d}
             \sum_{\vv k', \vv\ell' \in [s - r]_d}
             \delta_{\vv k + \vv k', \vv i}
             \delta_{\vv\ell + \vv\ell', \vv j}
             \sum_{\substack{\sigma_1 \in B(r, \vv k) \\ \tau_1 \in B(r, \vv\ell)}}
             \ketbra{\sigma_1}{\tau_1}
             {\underbrace{\sum_{\substack{\sigma_2 \in B(s - r, \vv k') \\ \tau_2 \in B(s - r, \vv\ell')}}
                \braket{\sigma_2 | \tau_2}}_{=\,\binom{s - r}{\vv k'} \delta_{\vv k', \vv\ell'}}}\text,
      \end{align*}
      where the multinomial coefficient was defined in \cref{eqn:defmulti}.
      We now insert the partial trace into $\rho$.
      Also note that $\vv k' \in [s - r]_d$ is a redundant condition that is already enforced by the multinomial coefficient.
      \begin{align*}
         \tr_2 \rho
         & = \sum_{a_1, a_2 = 0}^1
             \sum_{\vv i, \vv j \in [s]_d}
             \frac{\rho_{a_1, a_2}(\vv i, \vv j)}
                  {\sqrt{\binom{s}{\vv i} \binom{s}{\vv j}}}
             \sum_{\vv k, \vv\ell \in [r]_d}
             \sum_{\vv k' \in \Z^d}
             \delta_{\vv k + \vv k', \vv i}
             \delta_{\vv\ell + \vv k', \vv j}
             \binom{s - r}{\vv k'}
             \sum_{\substack{\sigma_1 \in B(r, \vv k) \\ \tau_1 \in B(r, \vv\ell)}}
             \ketbra{a_1 \sigma_1}{a_2 \tau_1} \\
         & = \sum_{a_1, a_2 = 0}^1
             \sum_{\vv i, \vv j \in [s]_d}
             \frac{\rho_{a_1, a_2}(\vv i, \vv j)}
                  {\sqrt{\binom{s}{\vv i} \binom{s}{\vv j}}}
             \sum_{\vv k, \vv\ell \in [r]_d}
             \sum_{\vv k' \in \Z^d}
             \delta_{\vv k + \vv k', \vv i}
             \delta_{\vv\ell + \vv k', \vv j}
             \binom{s - r}{\vv k'}
             \sqrt{\binom{r}{\vv k} \binom{r}{\vv\ell}}
             \ketbra{a_1 D^r_{\vv k}}{a_2 D^r_{\vv\ell}}
         \intertext{Rearranging sums and removing $\delta$s gives}
         \tr_2 \rho
         & = \sum_{a_1, a_2 = 0}^1
             \sum_{\vv k, \vv\ell \in [r]_d}
             \sqrt{\binom{r}{\vv k} \binom{r}{\vv\ell}}
             \ketbra{a_1 D^r_{\vv k}}{a_2 D^r_{\vv\ell}}
             \sum_{\vv i, \vv j \in [s]_d}
                \delta_{\vv\ell + \vv i - \vv k, \vv j}
                \frac{\rho_{a_1, a_2}(\vv i, \vv j)}
                     {\sqrt{\binom{s}{\vv i} \binom{s}{\vv j}}}
                \binom{s - r}{\vv i - \vv k} \\
         & = \sum_{a_1, a_2 = 0}^1
             \sum_{\vv k, \vv\ell \in [r]_d}
             \sqrt{\binom{r}{\vv k} \binom{r}{\vv\ell}}
             \ketbra{a_1 D^r_{\vv k}}{a_2 D^r_{\vv\ell}}
             \sum_{\vv i \in [s]_d}
                \mathbf1_{[s]_d}(\vv\ell + \vv i - \vv k)
                \frac{\rho_{a_1, a_2}(\vv i, \vv\ell + \vv i - \vv k)}
                     {\sqrt{\binom{s}{\vv i} \binom{s}{\vv\ell + \vv i - \vv k}}}
                \binom{s - r}{\vv i - \vv k} \\
         & = \sum_{a_1, a_2 = 0}^1
             \sum_{\vv k, \vv\ell \in [r]_d}
             \sqrt{\binom{r}{\vv k} \binom{r}{\vv\ell}}
             \ketbra{a_1 D^r_{\vv k}}{a_2 D^r_{\vv\ell}}
             \sum_{\vv i \in [s]_d - \vv k}
                \mathbf1_{[s]_d}(\vv\ell + \vv i)
                \frac{\rho_{a_1, a_2}(\vv i, \vv i + \vv\ell)}
                     {\sqrt{\binom{s}{\vv i} \binom{s}{\vv i + \vv\ell}}}
                \binom{s - r}{\vv i}\text.
      \end{align*}
      Here,
      \begin{equation*}
         \mathbf1_M(x)
         \coloneqq \begin{dcases}
            1 & \text{if } x \in M \\
            0 & \text{else}
         \end{dcases}
      \end{equation*}
      is the indicator function.
      All reformulations are valid under the convention $\frac00 = 0$.
      Note that addends with negative components in $\vv i$ are ruled out by the multinomial coefficient; and since $\vv k \in [r]_d$, subtracting $\vv k$ from a vector in $[s]_d$ will give---apart from those with negative contributions---a vector in $[s - r]_d$.
      In turn, $\vv\ell + \vv i$ will then automatically be an element of $[s]_d$, which finally gives
      \begin{equation*}
         \tr_2 \rho
         = \sum_{a_1, a_2 = 0}^1
           \sum_{\vv k, \vv\ell \in [r]_d}
           \sqrt{\binom{r}{\vv k} \binom{r}{\vv\ell}}
           \ketbra{a_1 D^r_{\vv k}}{a_2 D^r_{\vv\ell}}
              \sum_{\vv i \in [s - r]_d}
                 \frac{\rho_{a_1, a_2}(\vv i + \vv k, \vv i + \vv\ell)}
                      {\sqrt{\binom{s}{\vv i + \vv k} \binom{s}{\vv i + \vv\ell}}}
                 \binom{s - r}{\vv i}\text.
         \tag*{$\square$}
      \end{equation*}

   \section{Analytic forms for states and measurements}\label{app:analytic}
      In this appendix, we will give some analytic forms that were derived from the numerical data of the optimal states and measurements, as well as a few interesting purely numerical results.
      All results are only exemplary; an immediate way to generate a variation is to flip the qubits at Alice's or Bob's site, or to replace all states $\ket{D^x_y}$ by $\ket{D^x_{x - y}}$.

      Unless indicated differently, all results correspond to the case of unit probability, $p_{\mathrm{dist}} = 1$.
      In general, the optimal states with lower success probability will continuously arise from them, but they will have many more nonzero coefficients for which we cannot give an analytic form.
      As can be seen by the discontinuous slopes in \cref{fig:fidsr}, there are bifurcation points at which another structural branch of the state may happen to give the optimal fidelity.

      Finally, note that strictly speaking, unless $F = 1$, all results only give lower bounds for the maximization.

      We will write $\mathbb P(\ket x) \equiv \ketbra xx$.
      \subsection{Qubits}
         \begin{itemize}
            \item \begin{subequations}
               If $s \geq 2r -1$ (high\hyp loss regime), we have
               \begin{equation*}
                  C(E)
                    = 2 \mathbb P\biggl(
                         \frac{\ket{0 D^r_1} + \ket{1 D^r_0}}{\sqrt2}
                      \biggr)\text.
               \end{equation*}
               The sent state is
               \begin{equation*}
                  \ket\psi = \sqrt{\frac{r}{r + s}} \ket{0 D^s_1} + \sqrt{\frac{s}{r + s}} \ket{1 D^s_0}
               \end{equation*}
               and the resulting fidelity is $\frac12 + \frac{r}{2s}$.
               \end{subequations}

               For $p \neq 1$, we find $C(E) = 2 \mathbb P\Bigl( \frac{\ket{0 D^r_1} + \alpha \ket{1 D^r_0} - \sqrt{1 - \alpha^2} \ket{1 D^r_2}}{\sqrt2}\Bigr)$ with a single free parameter $\alpha$.
            \item \begin{subequations}
               For $s -1 = r > 2$, we have
               \begin{equation*}
                  C(E)
                     = 2 \mathbb P\biggl(
                          \frac{\ket{0 D^r_2} + \ket{1 D^r_0}}{\sqrt2}
                       \biggr) +
                       2 \mathbb P\biggl(
                          \frac{\ket{0 D^r_1} + \ket{1 D^r_r}}{\sqrt2}
                       \biggr)\text.
               \end{equation*}
               The sent state is
               \begin{equation*}
                  \ket\psi
                  = \frac{\ket{0 D^s_2}}{\sqrt2} +
                    \sqrt{\frac{r -1}{2(r +1)}} \ket{1 D^s_0} +
                    \frac{1}{\sqrt{r +1}} \ket{1 D^s_s}
               \end{equation*}
               and the resulting fidelity is $1$.
               \end{subequations}
            \item \begin{subequations}
               For $s -2 = r > 4$, we have
               \begin{equation*}
                  C(E)
                  = 2 \mathbb P\biggl(
                       \frac{\ket{0 D^r_1} + \ket{1 D^r_{r -1}}}{\sqrt2}
                    \biggr) +
                    2 \mathbb P\biggl(
                       \frac{\ket{0 D^r_2}}{\sqrt2} +
                       y \ket{1 D^r_0} +
                       x \ket{1 D^r_r}
                    \biggr) %\\
                    +
                    2 \mathbb P\biggl(
                       x \ket{0 D^r_0} -
                       y \ket{0 D^r_r} +
                       \frac{\ket{1 D^r_{r -2}}}{\sqrt2}
                    \biggr)\text,
               \end{equation*}
               where $x = \sqrt{\frac{1}{r (r -1)}}$ and $y = \sqrt{\frac12 - x^2}$.
               The sent state is
               \begin{equation*}
                  \ket\psi
                  = \frac12\sqrt{1 + \frac2r} \bigl[
                       \ket{0 D^s_2} + \ket{1 D^s_{s -2}}
                    \bigr] +
                    \frac12\sqrt{1 - \frac2r} \bigl[
                       -\ket{0 D^s_s} + \ket{1 D^s_0}
                    \bigr]\text.
               \end{equation*}
               and the resulting fidelity is $1$ (note $1 + \frac2r = \frac sr$).
               \end{subequations}
            \item \begin{subequations}
               For $r = 3$, $s = 5$, we have
               \begin{equation*}
                  C(E)
                  = 2 \mathbb P\biggl(
                       \frac{\ket{0 D^r_1} + \ket{1 D^r_3}}{\sqrt2}
                    \biggr) +
                    2 \mathbb P\biggl(
                       \frac{\ket{0 D^r_0} + \ket{1 D^r_2}}{\sqrt2}
                    \biggr)\text.
               \end{equation*}
               The sent state is
               \begin{equation*}
                  \ket\psi =
                     \frac12 \ket{0 D^s_1} + \sqrt{\frac{7}{30}} \ket{0 D^s_2} +
                     \frac{1}{\sqrt6} \ket{1 D^s_3} + \sqrt{\frac{7}{20}} \ket{1 D^s_4}\text.
               \end{equation*}
               \end{subequations}
               The fidelity is $\frac45$.
            \item \begin{subequations}
               For $r = 5$, $s = 8$, we have
               \begin{equation*}
                  C(E)
                  = 2 \mathbb P\biggl(
                       \frac{\ket{0 D^r_1} + \ket{1 D^r_4}}{\sqrt2}
                    \biggr) +
                    2 \mathbb P\biggl(
                       \frac{\ket{0 D^r_2}}{\sqrt2} +
                       \frac{3\ket{1 D^r_0} + \ket{1 D^r_5}}{2\sqrt5}
                    \biggr) %\\
                    +
                    2 \mathbb P\biggl(
                       \frac{\ket{0 D^r_0} - 3\ket{0 D^r_5}}{2\sqrt5} +
                       \frac{\ket{1 D^r_3}}{\sqrt2}
                    \biggr)\text.
               \end{equation*}
               The sent state is
               \begin{equation*}
                  \ket\psi =
                     \sqrt{\frac{7}{15}} \ket{0 D^s_2} -
                     \frac{1}{\sqrt{30}} \ket{0 D^s_7} +
                     \frac{2}{\sqrt{15}} \ket{1 D^s_0} +
                     \sqrt{\frac{7}{30}} \ket{1 D^s_5}\text.
               \end{equation*}
               \end{subequations}
               The fidelity is $\frac{27}{32}$.
            \item \begin{subequations}
               For $r = 7$, $s = 10$, we have
               \begin{multline*}
                  C(E)
                  = 2 \mathbb P\biggl(
                       a \ket{0 D^r_0} - \sqrt{\frac12 - a^2} \ket{0 D^r_4} +
                       b \ket{1 D^r_2} + \sqrt{\frac12 - b^2} \ket{1 D^r_6}
                    \biggr) \\ +
                    2 \mathbb P\biggl(
                       \sqrt{\frac12 - b^2} \ket{0 D^r_1} - b \ket{0 D^r_5} +
                       \sqrt{\frac12 - a^2} \ket{1 D^r_3} + a \ket{1 D^r_7}
                    \biggr) \\ +
                    2 \mathbb P\biggl(
                       \sqrt{\frac12 - b^2} \ket{0 D^r_2} - b \ket{0 D^r_6} +
                       a \ket{1 D^r_4} + \sqrt{\frac12 - a^2} \ket{1 D^r_0}
                    \biggr) \\ +
                    2 \mathbb P\biggl(
                       -a \ket{0 D^r_3} + \sqrt{\frac12 - a^2} \ket{0 D^r_7} +
                       \sqrt{\frac12 - b^2} \ket{1 D^r_5} + b \ket{1 D^r_1}
                    \biggr)\text.
               \end{multline*}
               The sent state is
               \begin{equation*}
                  \ket\psi
                  = \frac{x}{\sqrt2} \bigl[ \ket{0 D^s_2} + \ket{1 D^s_8} \bigr] +
                    \frac{y}{\sqrt2} \bigl[ -\ket{0 D^s_6} + \ket{1 D^s_4} \bigr] +
                    \sqrt{\frac{1 - x^2 - y^2}{2}} \bigl[ \ket{0 D^s_{10}} + \ket{1 D^s_0} \bigr]\text.
               \end{equation*}
               \end{subequations}
               The fidelity is $0.97422$, $a \approx 0.305$, $b \approx 0.282$, $x \approx 0.691$, $y \approx 0.561$.
            \item \begin{subequations}
               For $r = 4$, $s = 6$, we find an interesting behavior. The optimal fidelity is no longer achievable by a symmetric state, in the sense that if $\ket\psi = \ket0 \ket{\phi_0} + \ket1 \ket{\phi_1}$, we find slight deviations from $\norm{\ket{\phi_0}} = \norm{\ket{\phi_1}} = \frac{1}{\sqrt2}$.

               We have
               \begin{multline*}
                  C(E)
                  = 2 \mathbb P\biggl( \frac{a \ket{0 D^r_1} +
                                       \sqrt{1 - a^2} \ket{0 D^r_4} -
                                       b \ket{1 D^r_0} +
                                       \sqrt{1 - b^2} \ket{1 D^r_3}}{\sqrt2}
                               \biggr) \\
                    +
                    v \mathbb P\biggl( \frac{\sqrt{1 - b^2} \ket{0 D^r_0} +
                                       b \ket{0 D^r_3} +
                                       \sqrt{v -1} \ket{1 D^r_2}}{\sqrt v} \biggr) %\\
                    +
                    (3 - v) \mathbb P\biggl( \frac{\sqrt{2 - v} \ket{0 D^r_2} +
                                             \sqrt{1 - a^2} \ket{1 D^r_1} -
                                             a \ket{1 D^r_4}}{\sqrt{3 - v}} \biggr)\text,
               \end{multline*}
               where $v \approx 2$, $a \approx 1$ and $b \approx \frac{1}{\sqrt2}$.

               The sent state is
               \begin{equation*}
                  \ket\psi = m \ket{0 D^s_1} + n \ket{0 D^s_4} + o \ket{1 D^s_0} + p \ket{1 D^s_3} + q \ket{1 D^s_6}\text,
               \end{equation*}
               where the prefactors must be determined numerically.
               \end{subequations}
         \end{itemize}
      \subsection{Qudits}
         \begin{itemize}
            \item \begin{subequations}
               For $d = 3$, $s = 3$, $r = 2$, we now find unit fidelity (instead of $83.\overline3\,\%$).
               We have
               \begin{equation*}
                  C(E)
                  = 2 \mathbb P\biggl(
                       \frac{\ket{0 D^2_{0,0,2}} - \ket{1 D^2_{1,1,0}}}{\sqrt2}
                    \biggr) +
                    2 \mathbb P\biggl(
                       \frac{\ket{0 D^2_{0,2,0}} - \ket{1 D^2_{1,0,1}}}{\sqrt2}
                    \biggr) +
                    2 \mathbb P\biggl(
                       \frac{\ket{0 D^2_{2,0,0}} - \ket{1 D^2_{0,1,1}}}{\sqrt2}
                    \biggr)\text.
               \end{equation*}
               The sent state is
               \begin{equation*}
                  \ket\psi
                  = \frac{\ket{0 D^3_{0,0,3}} + \ket{0 D^3_{0,3,0}} + \ket{0 D^3_{3,0,0}}}{\sqrt6} -
                    \frac{\ket{1 D^3_{1,1,1}}}{\sqrt2}\text.
               \end{equation*}
            \end{subequations}
            \item \begin{subequations}
               For $d = 3$, $s = 5$, $r = 3$, we now find $F = 89.1769\,\%$ (instead of $80\,\%$).
               We have
               \begin{multline*}
                  C(E)
                  = 2 \mathbb P\biggl(
                       \frac{\ket{0 D^3_{1,0,2}} - \ket{1 D^3_{0,1,2}}}{\sqrt2}
                    \biggr) +
                    2 \mathbb P\biggl(
                       \frac{\ket{0 D^3_{1,2,0}} - \ket{1 D^3_{0,2,1}} }{\sqrt2}
                    \biggr) +
                    2 \mathbb P\biggl(
                       \frac{\ket{0 D^3_{3,0,0}} - \ket{1 D^3_{1,1,1}}}{\sqrt2}
                    \biggr) \\ +
                    2 \mathbb P\biggl(
                       \frac{\ket{0 D^3_{0,0,3}} - \ket{0 D^3_{0,3,0}} + \ket{1 D^3_{2,0,1}} + \ket{1 D^3_{2,1,0}}}{2}
                    \biggr) +
                    2 \mathbb P\biggl(
                       \frac{\ket{0 D^3_{0,0,3}} + \ket{0 D^3_{0,3,0}} - \ket{1 D^3_{2,0,1}} + \ket{1 D^3_{2,1,0}}}{2}
                    \biggr)\text.
               \end{multline*}
               The sent state is
               \begin{equation*}
                  \ket\psi
                  = \alpha \bigl[ \ket{0 D^5_{1,0,4}} - \ket{0 D^5_{1,4,0}} \bigr] -
                    \beta \ket{0 D^5_{5,0,0}} +
                    \gamma \bigl[ -\ket{1 D^5_{0,1,4}} + \ket{1 D^5_{0,4,1}} \bigr] +
                    \delta \ket{1 D^5_{3,1,1}}\text,
               \end{equation*}
               where $\alpha \approx 0.43$, $\beta \approx 0.42$, $\gamma \approx 0.22$, $\delta \approx 0.60$.
               In fact, we can give analytical expressions in terms of the roots of cubic polynomials.
               Here, $R_n\bigl(P(x)\bigr)$ represents the $n$\textsuperscript{th} real root (in ascending order, one\hyp indexed) of the polynomial $P(x)$.
               The fidelity is $\frac{1}{20} \bigl[ 9 + R_3(x^3 - 8x^2 - 23x +138) \bigr]$, $\alpha^2 = R_3(1\,471\,632 x^3 - 479\,136 x^2 + 39\,917 x - 324)$, $\beta^2 = R_1(91\,977 x^3 - 85\,560 x^2 + 25\,025 x - 2250)$, $\gamma^2 = R_1(490\,544 x^3 - 159\,712 x^2 + 15\,543 x - 432)$, $\delta^2 = R_3(30\,659 x^3 - 23\,529 x^2 + 5400 x - 324)$.
            \end{subequations}
         \end{itemize}

   \section{Previous analysis of redundant parity encoding}\label{app:rpe}
      In \cref{sec:rpe}, we explained the redundant parity encoding and gave the probability of a successful correction (or, in the language of our protocols, distillation),
      \begin{equation}
         p_{\mathrm{dist}} = \bigl(1 - \bar p_{\mathrm{trans}}^m\bigr)^n -
         \bigl(1 - \bar p_{\mathrm{trans}}^m - p_{\mathrm{trans}}^m\bigr)^n\text,
         \tag{\ref{eqn:rpecorrect}}
      \end{equation}
      where $\bar p_{\mathrm{trans}} \coloneqq 1 - p_{\mathrm{trans}}$.
      We noted that the previous analysis in \cite{bib:Munro2012} was overly optimistic, which we now want to detail.
      Note that $p_{\mathrm{dist}}$ is called $P_{\mathrm f}$ and $p_{\mathrm{trans}}$ is called $P$ in~\cite{bib:Munro2012}; we will use our notation also when we quote.

      In \cite{bib:Munro2012}, the authors describe ``the probability of no photon arriving in the $m$ photon logical qubit is $p_{\mathrm{dist}} = \bar p_{\mathrm{trans}}^m$'' as well as ``with $n$ logical qubits, the probability we do not receive at least one logical qubit without error is $p_{\mathrm{dist}} = (1 - p_{\mathrm{trans}}^m)^n$''.
      Their supplemental material suggests that they first determine $m$ by requiring a certain error rate $p_{\mathrm{dist}} < (1 - p_{\mathrm{trans}})^m$, where $p_{\mathrm{dist}}$ and $p_{\mathrm{trans}}$ are given.
      \emph{After this}, they consider the requirement of receiving at least one logical qubit intact and correspondingly adjust $n$ so that $p_{\mathrm{dist}} < (1 - p_{\mathrm{trans}}^m)^n$ also matches the requirement; here, $p_{\mathrm{dist}}$, $p_{\mathrm{trans}}$, and also $m$ are given.

      We will illustrate this by reconstructing the first entry, $p_{\mathrm{trans}} = 0.82$, $m = 4$, $n = 12$, of their Table~1.
      First, they want to put the probability that no physical qubit arrives from a logical qubit below the threshold $p_{\mathrm{dist}} < 0.001$: $(1 - p_{\mathrm{trans}})^m < 0.001$. This gives $m \geq 4.02832$, which they roughly translate to $m = 4$.
      Then, they want to ensure that the probability not to receive at least one logical qubit completely is below the threshold: $(1 - p_{\mathrm{trans}}^m)^n < 0.001$, which then, using $m = 4$, gives $n \geq 11.4804$, i.e., $n = 12$.
      However, this completely ignores that now with twelve logical qubits, the probability to have at least one physical qubit from each logical qubit is actually far lower: when determining~$m$, the calculation was based on the assumption of having only a single logical qubit.

      If we use their values of $m$, $n$ and $p_{\mathrm{trans}}$ to correctly calculate the success probability with our \cref{eqn:rpecorrect}, we end up at $p_{\mathrm{dist}} = 0.986761$---i.e., the failure probability is more than a percent, not less than a thousandth.
      In \cref{fig:rpe}, we use our \cref{eqn:rpecorrect} to check when the requirement $p_{\mathrm{dist}} \leq 0.001$ first holds for the first couple values of~$n$.
      \begin{figure}[htbp]
         \centering
         \begin{tikzpicture}
            \begin{axis}[xlabel=$n$, ylabel=minimal $m$ such that $p_{\mathrm{dist}} < 0.001$, legend pos=outer north east, height=10cm, ymin=30, ytick={31, 35, 40, 45, 50, 55, 60}, yticklabels={$1$, $35$, $40$, $45$, $50$, $55$, $60$}]
               \addplot[only marks, mpc0, mark size=1pt] coordinates {%
                  (1,35) (2,39) (3,41) (4,42) (5,43) (6,44) (7,45) (8,46) (9,46) (10,47) (11,47) (12,48) (13,48) (14,49) (15,49) (16,49) (17,50) (18,50) (19,50) (20,50) (21,51) (22,51) (23,51) (24,51) (25,52) (26,52) (27,52) (28,52) (29,52) (30,52) (31,53) (32,53) (33,53) (34,53) %  following points have second coordinate = 1
                  (35,31) (36,31) (37,31) (38,31) (39,31) (40,31) (41,31) (42,31) (43,31) (44,31) (45,31) (46,31) (47,31) (48,31) (49,31) (50,31) (51,31) (52,31) (53,31) (54,31) (55,31) (56,31) (57,31) (58,31) (59,31) (60,31) (61,31) (62,31) (63,31) (64,31) (65,31) (66,31) (67,31) (68,31) (69,31) (70,31) (71,31) (72,31) (73,31) (74,31) (75,31) (76,31) (77,31) (78,31) (79,31) (80,31) (81,31) (82,31) (83,31) (84,31) (85,31) (86,31) (87,31) (88,31) (89,31) (90,31) (91,31) (92,31) (93,31) (94,31) (95,31) (96,31) (97,31) (98,31) (99,31) (100,31)%
               };
               \addlegendentry{$m \geq 1$}
               \addplot[only marks, mpc1, mark size=.5pt] coordinates {(1,35) (2,39) (3,41) (4,42) (5,43) (6,44) (7,45) (8,46) (9,46) (10,47) (11,47) (12,48) (13,48) (14,49) (15,49) (16,49) (17,50) (18,50) (19,50) (20,50) (21,51) (22,51) (23,51) (24,51) (25,52) (26,52) (27,52) (28,52) (29,52) (30,52) (31,53) (32,53) (33,53) (34,53) (35,53) (36,53) (37,54) (38,54) (39,54) (40,54) (41,54) (42,54) (43,54) (44,54) (45,54) (46,55) (47,55) (48,55) (49,55) (50,55) (51,55) (52,55) (53,55) (54,55) (55,55) (56,56) (57,56) (58,56) (59,56) (60,56) (61,56) (62,56) (63,56) (64,56) (65,56) (66,56) (67,56) (68,57) (69,57) (70,57) (71,57) (72,57) (73,57) (74,57) (75,57) (76,57) (77,57) (78,57) (79,57) (80,57) (81,57) (82,58) (83,58) (84,58) (85,58) (86,58) (87,58) (88,58) (89,58) (90,58) (91,58) (92,58) (93,58) (94,58) (95,58) (96,58) (97,58) (98,58) (99,58) (100,59)};
               \addlegendentry{$m \geq 2$}
               \coordinate (discont) at (axis description cs: 0, .1);
               \coordinate (discont2) at (axis description cs: 1, .1);
            \end{axis}
            \draw[double] (discont) ++(-2mm, -2mm) -- ++(4mm, 4mm)
               (discont2) ++(-2mm, -2mm) -- ++(4mm, 4mm);
         \end{tikzpicture}
         \caption{Minimal values of $m$ (size of a block) and $n$ (number of blocks) required in a redundant parity encoding such that $p_{\mathrm{dist}} < 0.001$, with $p_{\mathrm{trans}} = 0.82$.
            For $n \geq 35$, we can already satisfy the condition without any inner encoding at all ($m = 1$); if we disregard this possibility, the discontinuity vanishes in the plotted region.}
         \label{fig:rpe}
      \end{figure}
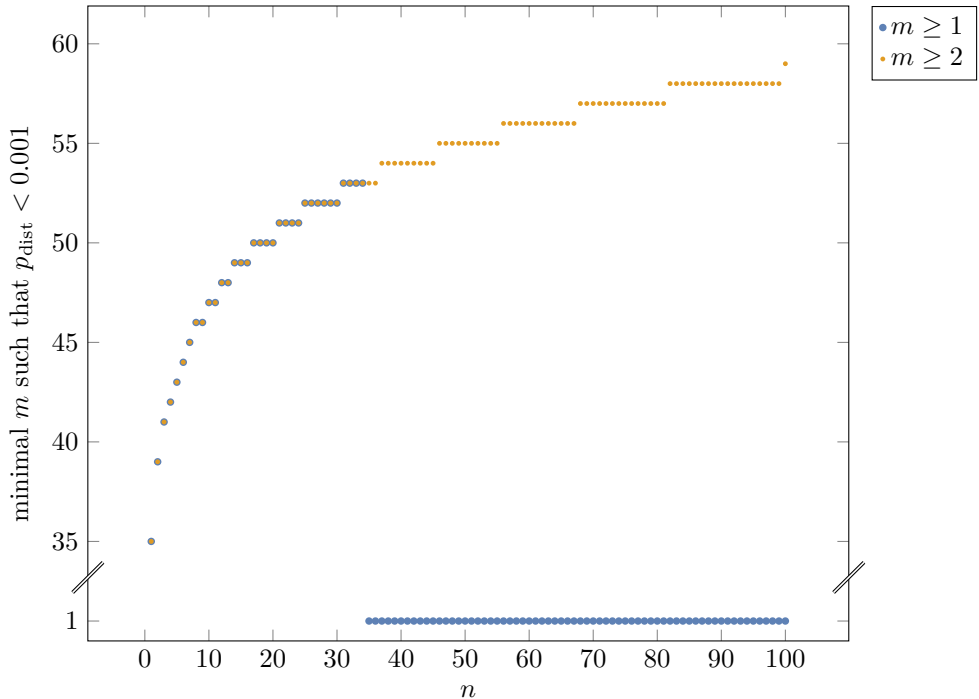
      \def\endwidetext{%
         \par
         \vskip6\p@
         \twocolumngrid\global\booltrue{@ignore}
         \booltrue{@endpe}
      }
   \end{widetext}
   \clearpage\leavevmode% Somehow the last page is missing if we don't do this
\end{document}